\documentclass[conference,a4paper,10pt,times]{IEEEtran}
\pdfoutput=1
\usepackage{amsmath,amsfonts,amssymb}
\usepackage{multirow}
\usepackage{xcolor}
\usepackage{enumitem}
\usepackage{algorithmicx}
\usepackage{algpseudocode}
\usepackage{algorithm}
\usepackage{subcaption}
\usepackage{mathtools, xparse} 
\usepackage{wrapfig}
\usepackage{xspace}
\usepackage[hyphens]{url}
\usepackage{hyperref}

\usepackage{graphicx}
\usepackage{subcaption}

\newcommand{\R}{\mathbb{R}}

\newcommand{\ourname}{\textsc{PRADA}\xspace}

\newcommand{\papernot}{\textsc{Papernot}\xspace}
\newcommand{\tramer}{\textsc{Tramer}\xspace}
\newcommand{\cvsearch}{\textsc{CV-search}\xspace}
\newcommand{\same}{\textsc{Same}\xspace}

\newcommand{\trnd}{\textsc{T-rnd}\xspace}
\newcommand{\colorr}{\textsc{Color}\xspace}

\newcommand{\RUagree}{\textsl{RU-agreement}\xspace}
\newcommand{\testagree}{\textsl{Test-agreement}\xspace}

\newcommand{\targeted}{\textsl{Targeted}\xspace}
\newcommand{\nontargeted}{\textsl{Non-targeted}\xspace}

\newif\iftechreport
\ifdefined\istechreport
\techreporttrue
\fi

\iftechreport
\newcommand\techreport[2]{\textcolor{blue}{TR: #1}}
\else
\newcommand\techreport[2]{{#2}}
\fi

\DeclareMathOperator*{\argmax}{arg\,max}

\begin{document}

\title{PRADA: Protecting Against DNN Model Stealing Attacks}

\author{
    \IEEEauthorblockN{Mika Juuti, Sebastian Szyller, Samuel Marchal, N. Asokan}
    \IEEEauthorblockA{Aalto University
    \\\{mika.juuti, sebastian.szyller, samuel.marchal\}@aalto.fi, asokan@acm.org}
}

\maketitle

\begin{abstract}
Machine learning (ML) applications are increasingly prevalent. Protecting the confidentiality of ML models becomes paramount for two reasons: (a) a model can be a business advantage to its owner, and (b) an adversary may use a stolen model to find \emph{transferable adversarial examples} that can evade classification by the original model. 
Access to the model can be restricted to be only via well-defined prediction APIs. 
Nevertheless, prediction APIs still provide enough information to allow an adversary to mount \emph{model extraction} attacks by sending repeated queries via the prediction API.

In this paper, we describe new model extraction attacks using novel approaches for generating synthetic queries, and optimizing training hyperparameters. Our attacks outperform state-of-the-art model extraction in terms of transferability of both targeted and non-targeted adversarial examples (up to +29-44 percentage points, pp), and 
prediction accuracy (up to +46 pp) on two datasets.
We provide take-aways on how to perform effective model extraction attacks. 

We then propose \ourname, the first step towards generic and effective detection of DNN model extraction attacks.
It analyzes the distribution of consecutive API queries and raises an alarm when this distribution deviates from benign behavior. We show that \ourname can detect all prior model extraction attacks with  no false positives.

\end{abstract}

\section{Introduction}
\label{sec:intro}

Recent advances in deep neural networks (DNN) have drastically improved the performance and reliability of machine learning (ML)-based decision making.
New business models like \textit{Machine-Learning-as-a-Service} (MLaaS) have emerged where the model itself is hosted in a secure cloud service, allowing clients to query the model via a cloud-based prediction API. 
ML models are also increasingly deployed on end-user devices, and can similarly be deployed behind APIs using hardware security mechanisms. 
Model owners can monetize their models by, e.g., having clients pay to use the prediction API. In these settings, the ML model represents business value underscoring the need to keep it confidential.

 
Increasing adoption of ML in various applications is also accompanied by an increase in attacks targeting ML-based systems a.k.a. \textit{adversarial machine learning}~\cite{papernot2016towards}. One such attack is forging \textit{adversarial examples}, which are samples specifically crafted to deceive a target ML model~\cite{goodfellow2014explaining}. 
To date, there are no effective defenses protecting against all such attacks~\cite{athalye2018obfuscated} but one partial mitigation is to protect the confidentiality of the ML model.


However, prediction APIs necessarily leak information.
This leakage of information is exploited by \textit{model extraction attacks}~\cite{Papernot:2017:PBA,tramer:2016:stealing} 
 where the adversary only has access to the prediction API of a target model which it can use as an oracle for returning predictions for the samples it submits. 
The adversary queries the target model iteratively using 
samples that are specifically crafted to maximize the extraction of information about the model internals via predictions returned by the model. 
The adversary uses this information to gradually train a \emph{substitute model}. The substitute model itself may be used in constructing future queries whose responses are used to further refine the substitute model. The goal of the adversary is to use the substitute model to (a) obtain predictions in the future, bypassing the original model, and thus depriving its owner of their business advantage, and/or (b) construct \emph{transferable adversarial examples}~\cite{vsrndic2014practical} that it can later use to deceive the original model into making incorrect predictions. The success of the adversary can thus be measured in terms of (a) prediction accuracy of the substitute model, and (b) transferability of adversarial samples obtained from the substitute model.


Prior extraction attacks are either narrowly scoped~\cite{Papernot:2017:PBA} (targeting transferability of a specific type of adversarial examples), or have been demonstrated only on simple models~\cite{tramer:2016:stealing}. 
We are not aware of any prior work describing effective generic techniques to detect/prevent DNN model extraction.

\textbf{Goal and contributions.}
Our goal is twofold. (1) demonstrate the \textit{feasibility of model extraction attack on DNN models}  
by proposing new, more effective attacks, and 
(2) develop an effective generic defense to model extraction.  By ``generic'', we mean applicability to models with any type of input data and any learning algorithm.
We claim the following contributions:
\begin{itemize}
  \itemsep0em
	\item \textbf{novel model extraction attacks} (Sect.~\ref{sec:approach}), which, unlike previous proposals, leverage the optimization of training hyperparameters and generalize synthetic data generation approaches. They outperform prior attacks in transferability of targeted and non-targeted adversarial examples (+29-44 pp) and prediction accuracy (up to +46 pp) (Sect.~\ref{sec:eval-end}). 
	\item \textbf{new insights on model extraction success factors} 
	showing that (a) cross-validated hyperparameter search outperforms selection of training hyperparameters (Sect.~\ref{sec:pre-eval}), 
	(b) prediction probabilities help improve transferability of adversarial examples, while class labels are sufficient for high
	prediction accuracy for the substitute model (Sect.~\ref{sec:seeds} -- \ref{sec:eval-end}),
	and (c) using the same architecture for the substitute model results in better transferablity while a more complex architecture
	can increase prediction accuracy (Sect.~\ref{sec:complexity}). 
	\item \textbf{a new technique, \ourname, to detect model extraction} which analyses the distribution of successive queries from a client and identifies deviations from a normal (Gaussian) distribution (Sect.~\ref{sec:detection-approach}). We show that it is \emph{effective}: 100\% detection rate and no false positives on all prior model extraction attacks (Sect.~\ref{sec:eval-detection}). To the best of our knowledge \ourname is the \emph{first generic technique} for detecting model extraction.
          
\end{itemize}
We share the source code for our attacks on request for research use. Our defense is available as open source\footnote{\url{https://github.com/SSGAalto/prada-protecting-against-dnn-model-stealing-attacks}}.



\section{Background}
\label{sec:method}

\subsection{Deep Neural Network (DNN)}
\label{sec:dnn_prelim}

A deep neural network (DNN) is a function $F(x)$ producing output $y \in \R ^m$ on input $x \in \R ^n$, where
$F(x)$ is a hierarchical composition of $k$ parametric functions $f_i$ ($i \in \lbrace1,k\rbrace$), each of which is a layer of neurons that apply \textit{activation functions} to the weighted output of the previous layer $f_{i-1}$. Each layer is parametrized by a weight matrix $\theta_i$, a bias $b_i$ and an activation function $\sigma_i$:  $f_i(x) = \sigma_i(\theta_i \cdot x + b_i)$ . Consequently a DNN can be formulated as follows:

\begin{equation}
	F(x) = f_k \circ f_{k-1} \circ \cdots \circ f_2 \circ f_1 \circ x
\end{equation}
\begin{equation}
	F(x) = \sigma_k(\theta_k \cdot \sigma_{k-1}(\theta_{k-1} \cdots \sigma_1(\theta_1 \cdot x + b_1) \cdots + b_{k-1}) + b_k)
\end{equation}

In this paper we focus on predictive DNNs used as $m$-class classifiers. The output of $F(x)$ is an $m$-dimensional vector containing the probabilities $p_j$ that $x$ belongs to each class $c_j$ for $j \in \lbrace1,m\rbrace$. 
The last activation function $\sigma_k$ is typically \textit{softmax}. 
A final prediction class\footnote{We use the hat notation \^~ to denote predictions} $\hat{F}$ is obtained by applying the \textit{argmax} function: $\hat{F}(x) = argmax(F(x)) = c$.

\subsection{Adversary Description}
\label{sec:adversary}

The adversary's objective is to ``steal'' a target machine learning model $F$ by making a series of prediction requests $U = \{x_1 , \dots, x_n \}$ to $F$. Responses $Y = \{\hat{F}(x_1) , \dots, \hat{F}(x_n) \}$ along with $U$, are used by adversary to train its substitute model $F'$. 
Model extraction attacks~\cite{Papernot:2017:PBA, tramer:2016:stealing} operate in a \emph{black-box} setting. The adversary does not have access 
to all target model internals, but has access to a prediction API.

Model extraction to date operates in settings where the adversary does not have a large set of ``natural'' samples.
These attacks require crafting and querying of \emph{synthetic samples} to extract information from $F$.
This attack pattern is increasingly more realistic given the emergence of the MLaaS paradigm, where one party (model provider) uses a private training set, domain expertise and computational power to train a model. The model is then made available to other parties (clients) via a prediction API on cloud-based platforms, e.g., AmazonML~\cite{amazonML} and AzureML~\cite{azureML}. The models are monetized by charging fees from clients for each prediction made by the models. 
The model provider may alternatively deploy models on client devices (e.g. smartphones or cameras), and rely on platform security mechanisms on these devices, to protect model confidentiality.
What is common to both scenarios is that while the models may be secured against physical theft, 
the prediction APIs will remain open, enabling model extraction attacks relying on predictions.

\subsection{Goals}
\label{sec:model_extraction}

Adversaries are incentivized to extract models for (Sect.~\ref{sec:intro}):
\begin{itemize}
  \itemsep0em
	\item \textbf{Reproduction of predictive behavior.} 
The purpose of the substitute model $F'$ is to reproduce as faithfully as possible the prediction of $F$ for a
known subspace $S$ of the whole input space $R ^n$, i.e. $\forall x \in S \subset \R ^n$. It may be:
\begin{itemize}
\item The whole space of input values $S = R^n$, in which case \emph{all} predictions made by $F'$ will match the predictions of $F$.
This \emph{Random Uniform Agreement} is measured by randomly sampling the input space. 
\item A relevant subset of the whole input space, e.g. all images $x$ that are in the subset ``digits'' when attacking a digit classifier.
\emph{Agreement} is measured by sampling a held-out test set that neither classifier has seen before. 
\end{itemize}

	\item \textbf{Transfer of adversarial examples.} 
Forging adversarial examples consists of finding minimal modifications $\epsilon$ for an input $x$ of class $c$ such that $x' = x + \epsilon$ is classified as $c' \neq c$ by a model $F$. 
Adversarial examples are either:
	\begin{itemize}
  \item \emph{Targeted,} where $x + \epsilon$ is created to change the classification of $x$ from $c$ to \emph{a specific class} $c'$. 
  \item \emph{Non-targeted,} where $x + \epsilon$ is created to change the classification of $x$ from $c$ to \emph{any other class} $c'$. 
  \end{itemize}	

\end{itemize}

Secondarily, adversaries want to minimize the number of prediction queries to $F$ in order to 
 (1) avoid detection and prevention of the attack, 
 (2) limit the amount of money spent for predictions, in the case of MLaaS prediction APIs, 
 and to (3) minimize the number of natural samples required to query the model.

\subsection{Adversary Model}
\label{sec:adversary_model}

\noindent\textbf{Attack surface.} We consider any scenario where the target model is isolated from clients by some means. This can be a \textit{remote isolation} where the model is hosted on a server or \textit{local isolation} on a personal device (e.g. smartphone) or an autonomous system (e.g., self-driving car, autonomous drone). We assume local and remote isolation provide same confidentiality guarantees and physical access does not help the adversary to overcome the isolation. Such guarantees can be typically enforced by hardware assisted TEEs~\cite{DBLP:journals/ieeesp/EkbergKA14}.
Increasing availability of lightweight hardware enhanced for DNNs~\cite{movidius} and the rise of federated learning will push machine learning computation to the edge~\cite{KonecnyMYRSB16,NIPS2017_7029}. Local isolation will become increasingly adopted to protect these models.
 
\noindent\textbf{Capabilities.} The adversary has black-box access to the isolated target model. 
It knows the shape of the input ($n$) and output ($m$) layers of the model. 
It knows the model architecture: intermediate layer shapes and activation function types. 
It can query samples $x$ to be processed by the model and gets the output predictions.
Predictions may be \emph{labels only} $\hat{F}(x)$, or full set of \emph{probabilities} $F(x)$
which is a $m$-dimensional vector containing the probabilities $p_i$ that $x$ 
belongs to each class $c_i$ for $i \in \lbrace1,m\rbrace$.
Classes $c_i$ are meaningful to the adversary, e.g., they correspond to digits, vehicles types, prescription drugs, etc. Thus, the adversary can assume what the input to the model looks like even though it may not know the exact distribution of the data used to train the target model\footnote{This is similar to~\cite{goodfellow2014explaining} but differ from~\cite{tramer:2016:stealing} which assumes that the adversary has no information about the purpose of classification, i.e, no intuition about what the input $x$ must look like. We consider that classes must be meaningful to provide utility to a client and that the adversary
  has access to a few natural samples for each class.}. 

\subsection{General Model Extraction Process}
\label{sec:pipeline}

We present a general process for extracting neural network models through prediction APIs.
Assuming a target model $F$, we want to learn a substitute model $F'$ to mimic behavior of $F$ (Section~\ref{sec:model_extraction}). 
The maximum number of queries may be limited to a query budget $b$.
We detail this model extraction process in Algorithm~\ref{alg:extraction} and 
discuss some of the steps next:

\begin{enumerate}
	\item [] \textbf{Initial data collection.} The adversary composes an initial set $U$ of unlabeled samples (\emph{seed samples}, row 6). The source of these samples is determined by the adversary's capabilities. Typically knowledge of input shape is assumed. 
	All samples are queried from $F$, and responses 
	are collected into dataset $L = \{U, \hat{F}(U) \}$ (row 7). 
	\item [] \textbf{Architecture and hyperparameters.} The adversary selects a neural network architecture (row 8) and hyperparameters (row 9) for $F'$. 
$F'$ is trained with $L$. After this, $\rho$ \textit{duplication rounds} (iterative steps) are run. 
	\item [] \textbf{Duplication rounds.} The adversary increases coverage of the input space by generating synthetic samples. This generation typically leverages knowledge of $F$ acquired until then: labeled training samples $L$ and current $F'$. This synthetic data is allocated to a new set $U$ (row 13). 
	All, or part, of the unlabeled synthetic samples $x \in U$ are queried from $F$, to get predictions $\hat{F}(x)$. These new labeled samples 
	are added to $L$
	 (row 14).
	Labeled samples $L$ are used for training $F'$ (row 15).  
\end{enumerate}

\textit{Duplication rounds} are repeated until the prediction query budget $b$ is consumed or termination occurs otherwise. The outcome is a substitute model $F'$ that mimics behavior of $F$.

\begin{algorithm}
\caption{Model extraction process with the goal of extracting classifier $F$, 
given initial unlabeled seed samples $X$ and a substitute model $F'$ (initially random). 
 }
\label{euclid}
\begin{algorithmic}[1]
\Procedure{Label}{$\{x_1, \dots, x_n \}, F$}
\State \textbf{return} $\{\hat{F}(x_1), \dots , \hat{F}(x_n) \}$\Comment{Return predictions}
\EndProcedure
\State 
\Procedure{ExtractModel}{$F$}
\State ${U} \gets \textit{Initial data collection}$
\State ${L} \gets \{U, ~\textproc{Label}(U, F) \}$
\State $F' \gets$ \textit{Select architecture}
\State ${H} \gets$ \textit{Resolve hyperparameters}\Comment{cf. Sec.~\ref{sec:hyperparameters}}
\State $F' \gets \textproc{Initialize}(F')$\Comment{Set random weights}
\State $F' \gets \textproc{Train}(F'~|~{L}, {H})$
\For{$i\gets 1, \rho$}\Comment{$\rho$ \textit{duplication rounds}}
\State $U \gets $ \textit{Create synthetic samples}\Comment{cf. Sec.~\ref{sec:synthetic_generation}}
\State $L \gets \{ ~{L} \cup \{U, ~\textproc{Label}(U, F) \} ~\}$
\State $F' \gets \textproc{Train}(F'~|~{L}, {H})$
\EndFor
\State \textbf{return} $F'$
\EndProcedure
\end{algorithmic}
\label{alg:extraction}
\end{algorithm}

\subsection{Prior Model Extraction Attacks}
\label{sec:prior}

We present two main techniques that have been introduced to date for model extraction. These are used as a baseline to 
improve model extraction attacks and compare performance 
(Sect.~\ref{sec:eval}) and to evaluate our detection approach (Sect.~\ref{sec:detection}).

\subsubsection{\tramer attack~\cite{tramer:2016:stealing}}
Tramer et al. introduced several attacks to extract simple ML models including the one-layer logistic regression model with an \textit{equation solving} attack and decision trees with a \textit{path finding} attack. Both have high efficiency and require a few prediction queries but are limited to the simple models mentioned. These attacks are specifically designed to reach very high \textit{Random Uniform Agreement} (Sect.~\ref{sec:model_extraction}).
The authors also introduce an extraction method targeting shallow neural networks that we present below according to our process (Sect.~\ref{sec:pipeline}).

Initial data consists of a set $U$ of uniformly selected random points of the input space (row 6). No natural samples are used. 
The attack assumes knowledge of the model architecture, hyperparameters and training strategy used for $F$ (rows 7--8).  

The main contribution for extracting neural networks lies in the prediction queries (row 13), where they introduce three strategies for querying 
additional data points from $F$. 
The first selects these samples randomly. The second called \textit{line-search retraining} selects new points closest to the decision boundary of the current $F'$ using a line search technique. 
The last is \textit{adaptive retraining} which has same intuition of querying samples close to the decision boundary, but it employs active learning techniques~\cite{cohn1994improving}. 
The first two techniques are implemented\footnote{https://github.com/ftramer/Steal-ML}, and we evaluate the strictly stronger line-search retraining technique as \tramer in this work. 
This technique initially queries 25\% of the budget with random data (row 6), and then constructs line-search queries with 75\% of remaining budget in one duplication round (row 13).


\subsubsection{\papernot attack~\cite{Papernot:2017:PBA}}
Papernot et al. introduced a model extraction attack that is specifically designed at forging transferable \textit{non-targeted} adversarial examples (Sect.~\ref{sec:model_extraction}). We present this technique according to our process.

Initial data consists of a small set of natural samples (row 6). These are disjoint samples  
but distributed similarly as the target model's training data. 
Seed samples are balanced (same number of samples per class) and their required number increases with the model input dimensionality.
The attack does not assume knowledge of $F$, hyperparameters or training strategy, however  
expert knowledge is used to select a model architecture ``appropriate'' for the classification task of $F$ (row 8). 
Two strategies are proposed to query $F$ (row 14), one queries the whole set $U$ while the other called \textit{reservoir sampling} queries a random subset of $X\%$ samples from $U$. Unselected samples are thrown away.
\papernot is defined with a fixed training strategy: Stochastic Gradient Descent~\cite{murphy2012machine} with learning rate $0.01$, and momentum $0.9$. 
$F'$ is trained for a very short time (10 epochs) at each duplication round, to save time and to avoid overfitting $L$.

They introduce the \textit{Jacobian-based Dataset Augmentation} (JbDA) technique for generating synthetic samples (row 13). 
It relies on computing the Jacobian matrices with the current $F'$ evaluated on the already labeled samples in $L$. 
Each element $x \in L$ is modified by adding the sign of the Jacobian matrix $\nabla_x \mathcal{L} (F'(x,c_i))$ dimension corresponding to the label assigned to $x$ by $F$, evaluated with regards to the classification loss $\mathcal{L}$. 
Thus, the set  $U$ is extended with $\lbrace x + \lambda \cdot sign(\nabla_x \mathcal{L} (F'(x,c_i) ) ) \rbrace,~\forall x \in L$. 
$U$ has the same size as $L$, which means that the number of generated synthetic samples doubles at each iteration.
$\lambda$ is fixed to 25.5/255 in their evaluation.
Thus, the creation of synthetic samples is identical to calculation of adversarial examples using the \emph{Fast Gradient Sign Method} (FGSM)~\cite{goodfellow2014explaining} on $F'$, and augmenting the attacker's set $L$ with their classification labels $\hat{F}(x)$. The duplication rounds are repeated for a predefined number of $\rho$ iterations, which they call \textit{substitute training epochs}. 

\section{DNN Model Extraction Framework}
\label{sec:approach}

Techniques proposed to date~\cite{tramer:2016:stealing,Papernot:2017:PBA} are narrowly scoped and explored solutions for only some of the required steps (Sect.~\ref{sec:model_extraction}). 
We investigate several strategies for two crucial steps of the model extraction process: \textit{selecting hyperparameters} (Algorithm~\ref{alg:extraction}, row 9) and \textit{synthetic sample generation} (row 13), and investigate what advantage probabilities (rather than label responses) give to the adversary (rows 7 and 14). 
In addition, in Sect.~\ref{sec:seeds} we explore the impact of natural sample availability during initial data collection (row 6), and in Sect.~\ref{sec:complexity} what impact mismatch in model architectures have on attack performance (row 8).

\subsection{Hyperparameters}
\label{sec:hyperparameters}

Preditive performance of neural networks is highly dependant on hyperparameters used for training. 
These include the \emph{learning rate},  and the \emph{number of training epochs}. 
Too low a learning rate may preclude finding the optimal solution before termination
whereas too high a rate can overshoot optimal solutions.
There are essentially three ways of choosing hyperparameters in model extraction attacks:
\begin{itemize}
\item \emph{Rule-of-thumb.} Use some heuristic. 
E.g. \papernot~\cite{Papernot:2017:PBA} uses a fixed learning rate and small number of epochs.
\item \same. Copy from the target model. 
This may be obtained via insider knowledge, or through state-of-the-art attacks~\cite{joon2018towards}.
\item \cvsearch. Do a \emph{cross-validation search}
on the initial seed samples (row 6). 
\end{itemize}

In this paper, we conduct \cvsearch by five-fold cross-validation. 
Five-fold cross-validation proceeds as follows.
For each hyperparameter combination we want to test out, the initial labeled 
dataset $L$ (row 7) is divided into 5 non-overlapping sets. 
The average accuracy is aggregated over the sets, is saved, 
and next hyperparameter combination is tested out. 
The process is repeated 5 times, each with a different validation set. 
The hyperparameter combination
that produces the best accuracy on validation sets is selected for the rest of the attack. 

Given a finite time, not all parameter combinations can be tested out. While strategies
like \emph{grid search} and \emph{random search} \cite{goodfellow2016deep} are popular, 
Bayesian hyperparameter optimization~\cite{snoek2012practical} is more 
efficient: after first querying
some initial samples, it
estimates what validation 
accuracies certain hyperparameter
combinations might have,
along with the uncertainty of these estimates.
Then the next test hyperparameter combinations are chosen as
the ones that have either high expected value or high 
uncertainty\footnote{\url{https://github.com/fmfn/BayesianOptimization}}.
\cvsearch is done with dropout training~\cite{murphy2012machine}.
We detail our \cvsearch procedure using Bayesian Optimization in Algorithm~\ref{alg:cv-search}. 
Learning rate is searched between $10^{-4}$ and $10^{-2}$, and training epochs between 10 and 320.
Both are searched in log-scale. 


\begin{algorithm}
\caption{Five-fold cross-validation (CV) search using bayesian optimization, 
given labeled dataset $L$,
and closed linear span of $\mathcal{H}$ (hyperparameters range:  [learning rate] $\times$ [train epochs]). The procedure searches for the best
hyperparameter combination $H_{i^*}$ that maximizes 5-fold CV accuracy.}
\label{euclid}
\begin{algorithmic}[1]
\Procedure{Sample}{$L_{\rm train}, L_{\rm val}, H$}\Comment{Calc. CV-accuracy}
\State $F' \gets \textproc{Initialize}(F')$\Comment{Set random weights}
\State $F' \gets \textproc{Train}(F'~|~L_{\rm train}, {H})$
\State \textsl{accuracy} $\gets \textproc{Evaluate}(F', L_{\rm val})$
\State \textbf{return} \textsl{accuracy}
\EndProcedure
\State
\Procedure{5-FoldSample}{$H$}\Comment{Average over 5 folds}
\For{$i\gets 1, 5$}
\State $\textsl{acc}_i \gets \textproc{Sample}(L_{\rm train}^i, L_{\rm val}^i, H )$
\EndFor
\State \textbf{return} \textproc{Mean}$(\textsl{acc}_1 , \dots, \textsl{acc}_5 )$
\EndProcedure
\State
\Procedure{CV-Search}{$F', L, \mathcal{H}$}
\State $(L_{\rm train}^1, L_{\rm val}^1), \dots, (L_{\rm train}^5, L_{\rm val}^5) \gets \textproc{KFolds}(L,5)$
\For{$i\gets 1, 4$}\Comment{Sample each corner of $\mathcal{H}$}
  \State ${H}_{\rm i} \gets \textsl{GetVertex} (\mathcal{H}, i)$
  \State $y_{\rm i} \gets \textproc{5-FoldSample}({H}_{\rm i})$
\EndFor
\For{$i\gets 5, 15$}\Comment{Sample randomly inside $\mathcal{H}$}
  \State ${H}_{\rm i} \gets \textsl{UniformRandom} (\mathcal{H})$
  \State $y_{\rm i} \gets \textproc{5-FoldSample}({H}_{\rm i})$
\EndFor
\For{$i\gets 16, 30$}\Comment{Sample with Gauss. Process \textsl{GP}}
\State $\textsl{GP} \gets \textproc{Initialize}()$\Comment{Set random weights}
\State \textit{GP} $ \gets$ \textit{Train GP to predict $y_{1, \dots, i-1}$ from $H_{1, \dots, i-1}$}
\State ${H}_{\rm i} \gets $ \textit{Find next value that GP perceives maximizes ``expected value + standard deviation''}
\State $y_{\rm i} \gets \textproc{5-FoldSample}({H}_{\rm i})$
\EndFor
\State $i^* \gets \argmax({y}_{\rm i})$
\State \textbf{return} $H_{i^*}$
\EndProcedure
\end{algorithmic}
\label{alg:cv-search}
\end{algorithm}

\subsection{Adversarial Example Crafting}
\label{sec:adversarial_example}

Adversarial examples\techreport{, both targeted and non-targeted,}{} are crafted by modifying 
samples $x \in R ^n$ with the Jacobian matrix for a given DNN $F$ with $C$ classes, which in
turn tells what the impact of each feature is on the overall classification loss $\mathcal{L}$ \cite{murphy2012machine}. 

The Jacobian is used for finding out \emph{how} to modify the
features of a sample $x$ such that the sample is classified as something different
from its genuine class $c_i$.
To modify $x$ into a \emph{targeted} adversarial example of class $c_j \neq c_i$,
 the Jacobian component on column $j$ is used, such that $x$ is modified in the
 \emph{negative} gradient direction $x' \leftarrow x - f(\nabla_x \mathcal{L} (F(x,c_j) ) )$ with some function 
\techreport{$f$, since modifying it in this direction increases the likelihood of classifying it as
 a member of class $c_j$. }{$f$.} 
 
To create a \emph{non-targeted} adversarial example, the $i$th column of the 
Jacobian is used. The sample $x$ is modified in the general \emph{positive} 
gradient direction, to \emph{decrease} the likelihood of classifying it
as a member of class $c_i$: $x' \leftarrow x + f(\nabla_x \mathcal{L} (F(x,c_i) ) )$
 For brevity we only discuss the non-targeted variant here.

The form of $f$ determines the adversarial example crafting algorithm. Popular
choices are Fast Gradient Sign Method \emph{FGSM}~\cite{goodfellow2014explaining}, 
and its iterative variants \emph{I-FGSM}~\cite{kurakin2016adversarial} and 
\emph{MI-FGSM}~\cite{dong2017boosting}. The overall modification for each of these
algorithms is bounded to remain within an $L_\infty$ distance of $\epsilon$. 

\paragraph{FGSM} A sample $x$ of class $c$ is modified by the \emph{sign}-function of the gradient and multiplied by a small $\epsilon$,

\begin{equation}
x' \leftarrow x + \epsilon \cdot sign(\nabla_x \mathcal{L} (F(x,c_i) ) )
\end{equation}

FGSM is called a ``one-step method'', and until recently, it was thought that
 these methods are most effective at producing transferable adversarial examples~\cite{dong2017boosting}. 

\paragraph{I-FGSM} Iterative FGSM subdivides modifications into $k$ steps,
such that every iterative modification is done with FGSM with step size $\frac{\epsilon}{k}$. 

\techreport{I-FGSM has been demonstrated to be strictly superior to FGSM in producing adversarial examples
for white-box DNNs, i.e. adversarial examples can be produced with smaller modifications $\epsilon$. However, the method has been shown to have difficulties in producing transferable adversarial examples~\cite{dong2017boosting}.
}{}

\paragraph{MI-FGSM} Momentum Iterative FGSM was recently shown to be the strongest method
of creating transferable adversarial examples when attacking DNN models~\cite{dong2017boosting}. 
MI-FGSM includes a momentum term that accumulates previous gradient directions~\cite{dong2017boosting}. 
MI-FGSM won both the targeted and non-targeted adversarial example challenge at NIPS 2017 Adversarial Attack competition. 

\begin{figure}
    \centering
    \begin{subfigure}[b]{0.22\textwidth}
        \includegraphics[scale=.32]{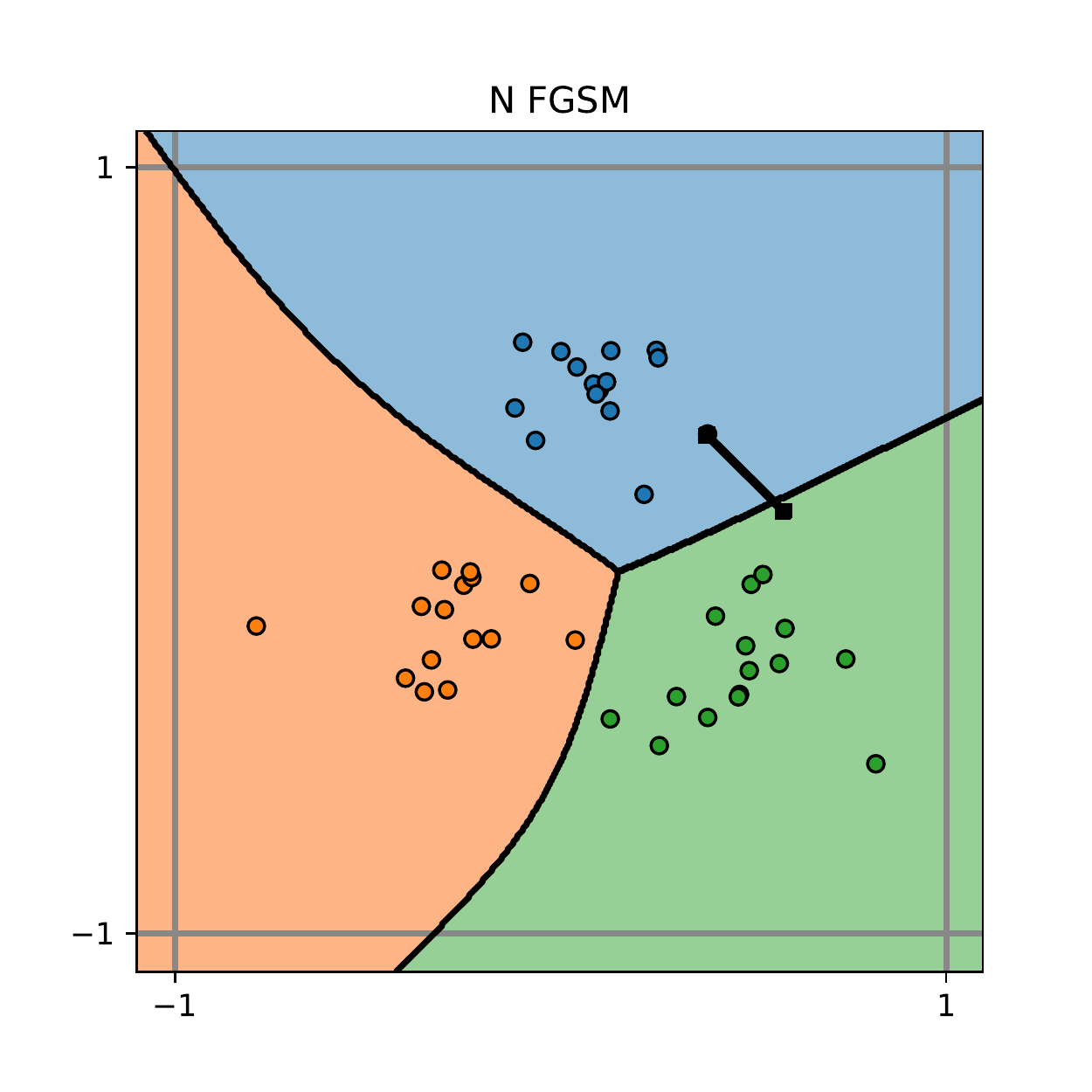}
        \caption{Non-targeted FGSM}
        \label{fig:fgsm_n}
    \end{subfigure}
    ~ 
    \begin{subfigure}[b]{0.22\textwidth}
        \includegraphics[scale=.32]{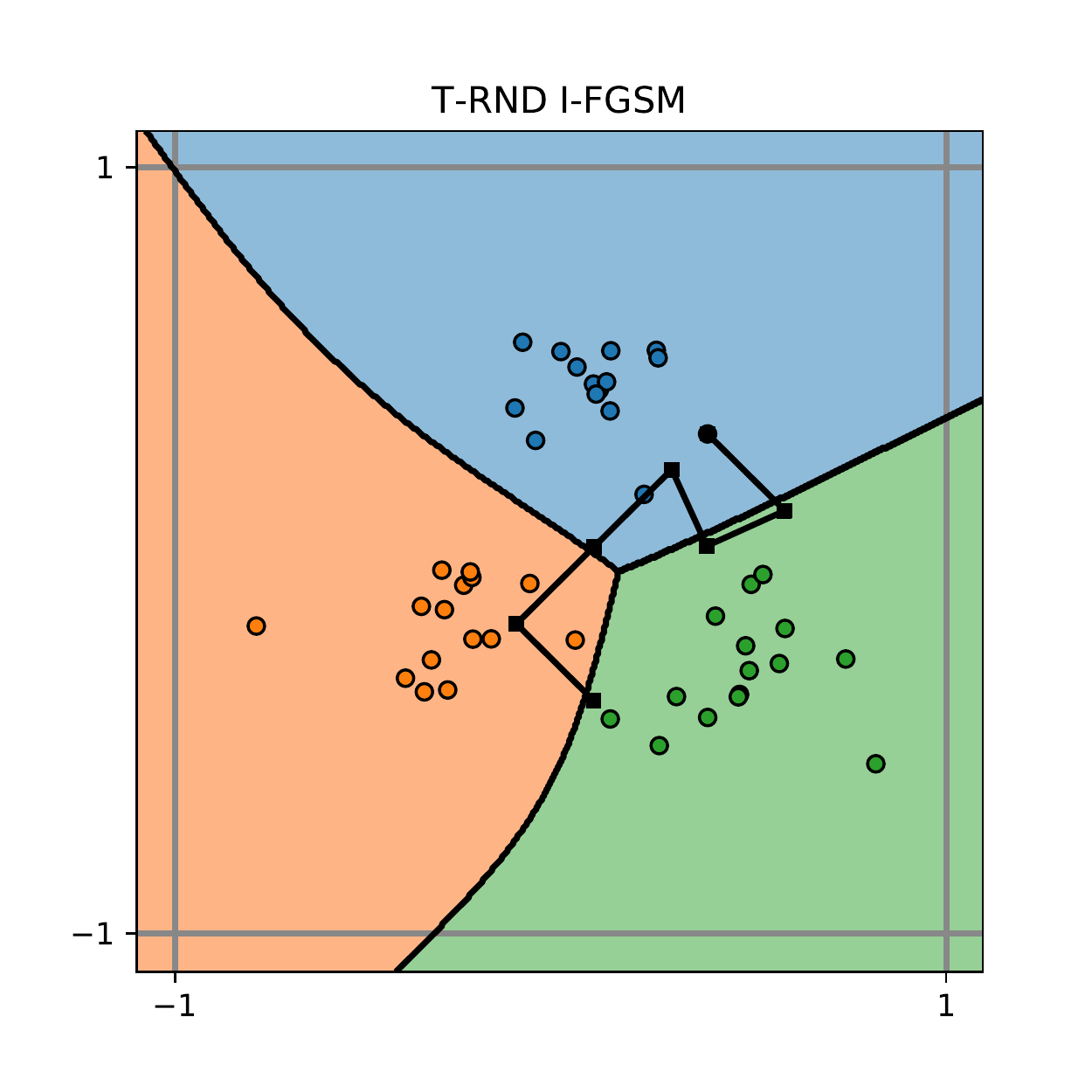}
        \caption{T-RND I-FGSM}
        \label{fig:I-FGSM_t}
    \end{subfigure}

    \caption{Synthetic sample generation against a multi-layer perceptron. We show \emph{six sequential steps}. Left: the non-targeted FGSM~\cite{Papernot:2017:PBA} does not generate novel data points after the first step. Right: T-RND I-FGSM avoids this by varying the contribution of features, and targeting random classes. }\label{fig:synthetics_schema}
\end{figure}

\subsection{Synthetic Sample Generation}
\label{sec:synthetic_generation}

Synthetic samples in model extraction attacks can be constructed either using the partially trained substitute model $F'$,
or independently of it. We call these strategies \emph{Jacobian-based Synthetic Sample Generation}
and \emph{Random Synthetic Sample Generation} respectively. 

We create new synthetic samples with regards to all previously labeled data:
the number of new samples increases exponentially with the number of duplication rounds $\rho$ (Algorithm~\ref{alg:extraction}, row 12). 
We call the rate at which
the number of synthetic samples grows the \emph{expansion factor} $k$. 



\subsubsection{Jacobian-based Synthetic Sample Generation}

These variants use adversarial example crafting algorithms (Sect.~\ref{sec:adversarial_example})
to produce new synthetic samples. 
Previous work~\cite{Papernot:2017:PBA} considered using non-targeted FGSM. 
We consider several choices, particularly targeted variants. 
All variants produce synthetic samples that step closer and closer to the perceived classification boundaries over the course of several duplication rounds (Algorithm~\ref{alg:extraction}, row 12). 

We illustrate the intuition for the effect different 
algorithms have in Fig.~\ref{fig:synthetics_schema}. For this, we trained a multi-layer 
perceptron (MLP)~\cite{murphy2012machine} over two-dimensional toy data with three classes. 
We show six steps, which corresponds to six duplication rounds in \cite{Papernot:2017:PBA}. 
We first demonstrate the synthetic sample crafting method of using \underline{N}on-targeted \underline{FGSM}~\cite{Papernot:2017:PBA}. 
Non-targeted variants try to greedily move towards the closest other class. 
If the classifier is not updated sufficiently between runs, the algorithm behaves like in 
Fig.~\ref{fig:fgsm_n}, where synthetic samples start to overlap,
and do not contribute with new information about the target model $F$. 
The overlapping behavior can be avoided to a certain degree by stepping in a
\emph{\underline{t}argeted} \underline{rand}omly chosen direction (T-RND), as in Figure~\ref{fig:I-FGSM_t}. Importantly, overlap can be further 
avoided by using \underline{i}terative \underline{FGSM} methods (I-FGSM) that can vary the contribution of different feature components. 



For non-targeted methods, the expansion factor is always $k=2$. 
However, for targeted variants, $k$ can be as high as the number of classes $C$. 
We set $k=4$ for the targeted variants in our tests.

\subsubsection{Random Synthetic Sample Generation}

In addition to these, we consider a generic synthetic sample generation
method: randomly perturbing \underline{color} channels (COLOR). 
For grayscale images, COLOR randomly increases
or decreases luminosity by a step size $\lambda$. For colored images, COLOR
 randomly perturbs the color channel of each pixel by the same amount for a 
given color channel. 
Random synthetic sample generation methods can have arbitrary expansion factors,
but we set $k=4$ in this paper. 

\section{DNN Model Extraction: Evaluation}
\label{sec:eval}

\techreport{
We investigate several properties of model extraction attacks and construct systematized tests to understand their effect.
First, we replicate existing techniques~\cite{tramer:2016:stealing,Papernot:2017:PBA} and 
evaluate hyperparameter choices made in~\cite{Papernot:2017:PBA}. We compare model extraction
performance using two hyperparameter choice strategies: \same and \cvsearch (Sect.~\ref{sec:hyperparameters}). 
We show that cross-validation search has a crucial impact on model extraction effectiveness, 
and can produce both competitive agreement and transferability already \emph{without synthetic samples}. 
We demonstrate that iterative methods of crafting adversarial examples 
gain significantly (2-3$\times$) benefits from model extraction attacks, compared to one-step methods. 

We then investigate the co-dependencies between the available number of seed samples, access to probabilities and extraction effectiveness.
We demonstrate that larger sets of seed samples significantly increase model extraction performance.
Then we evaluate the performance of our attacks and show they outperform the state-of-the-art. Finally we study the impact of the complexity of the target model on extraction effectiveness.
}{
In this section, we replicate prior techniques for model extraction~\cite{tramer:2016:stealing,Papernot:2017:PBA}, to explore the effect of different parameter
choices and develop new, more effective model extraction attacks.
}

\subsection{Experiment Setup}
\label{sec:exp-setup}

\paragraph{Datasets and target model description}

We evaluate two datasets: MNIST~\cite{lecun2010mnist} for digit recognition and GTSRB~\cite{stallkamp2011german} for traffic sign recognition. 
We chose these datasets because they had been evaluated in previous studies~\cite{Papernot:2017:PBA}, and we wish to validate their observations under our adversary model (Sect.~\ref{sec:model_extraction} -- \ref{sec:adversary_model}). 
MNIST contains 70,000 images of 28$\times$28 grayscale digits (10 classes). Its training set contains 60,000 and the rest are in the test set. GTSRB contains 39,209 images in the training set, and 12,630 images in the test set (43 classes). Images in GTSRB have different shapes (15$\times$15 to 215$\times$215); we normalize them to 32$\times$32. We additionally scale feature values for both datasets to the range [-1, 1].

%
%

\begin{wraptable}{r}{0.22\textwidth}
  \begin{center}
    \begin{tabular}{c c} \hline
    \textbf{MNIST} & \textbf{GTSRB} \\ \hline
    
     conv2-32 & conv2-64 \\ 
     maxpool2 & maxpool2 \\ \hline    
     conv2-64 & conv2-64 \\ 
     maxpool2 & maxpool2 \\  \hline
     FC-200 & FC-200\\ \hline
     FC-10 & FC-100 \\\hline
     & FC-43\\  \hline
    
    \end{tabular}
    \caption{Target models architecture (ReLU activation between blocks).}
    \label{table:architectures}
  \end{center}
  \vspace{-0.5cm}
\end{wraptable} 

\begin{figure*}[ht]
	\begin{subfigure}{0.23\textwidth}
	\centering
  \includegraphics[width=1.\columnwidth]{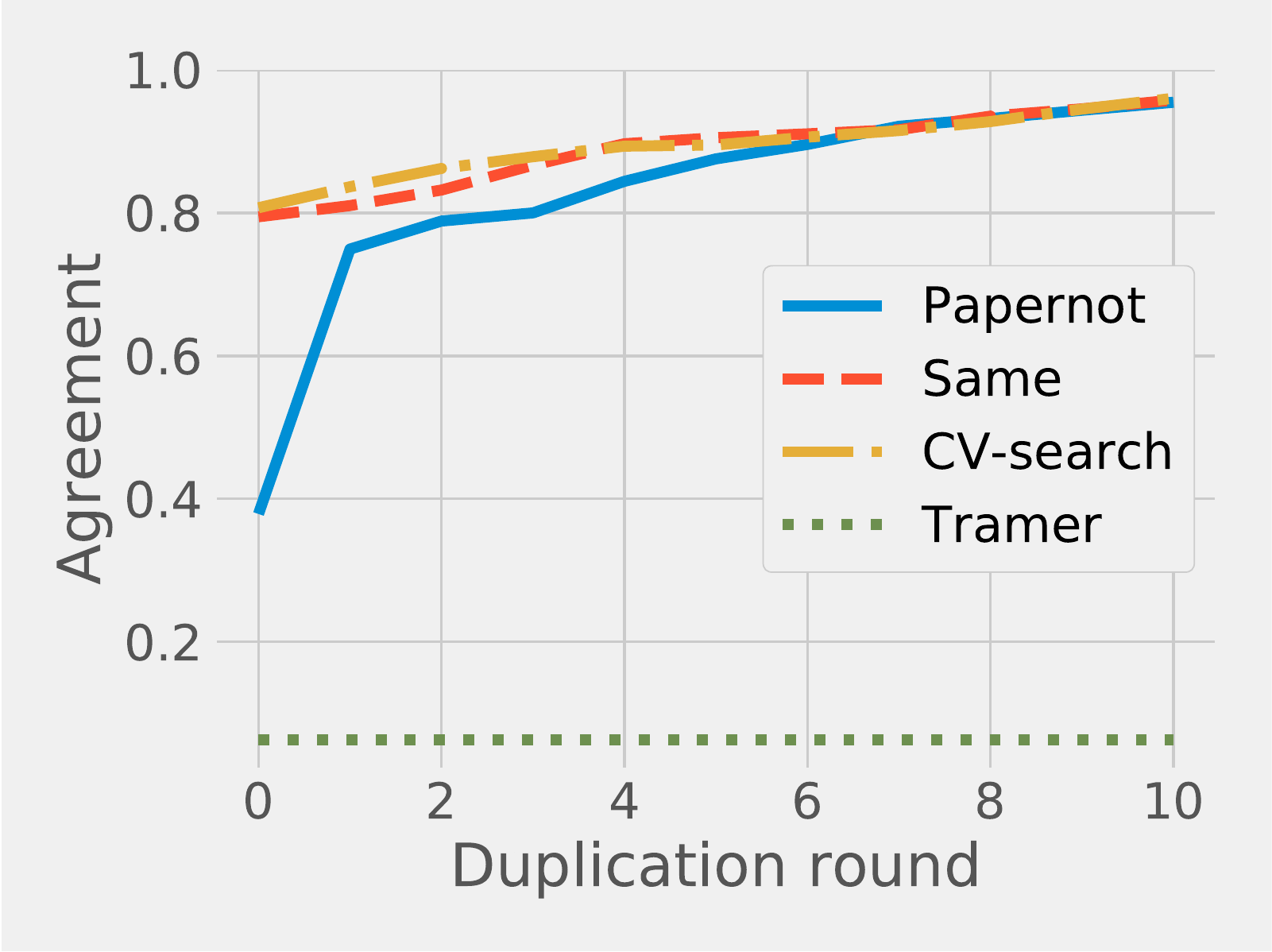}
		\caption{\testagree.}
	\end{subfigure}	
	\begin{subfigure}{0.23\textwidth}
	\centering
	\includegraphics[width=1.\columnwidth]{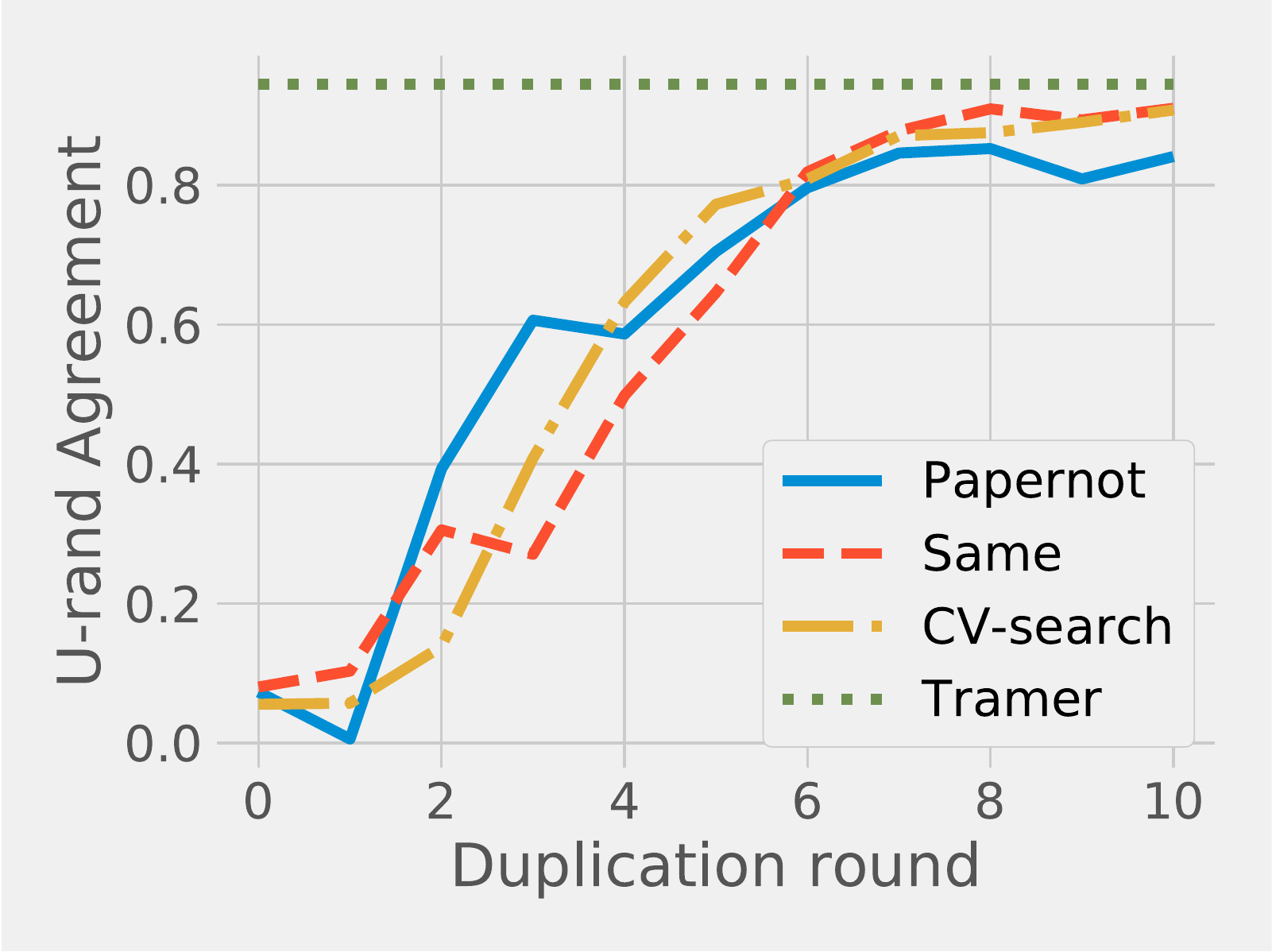}	
		\caption{\RUagree.}
	\end{subfigure}
      \hspace{0.3cm}
	\begin{subfigure}{0.23\textwidth}
	\centering
	\includegraphics[width=1.\columnwidth]{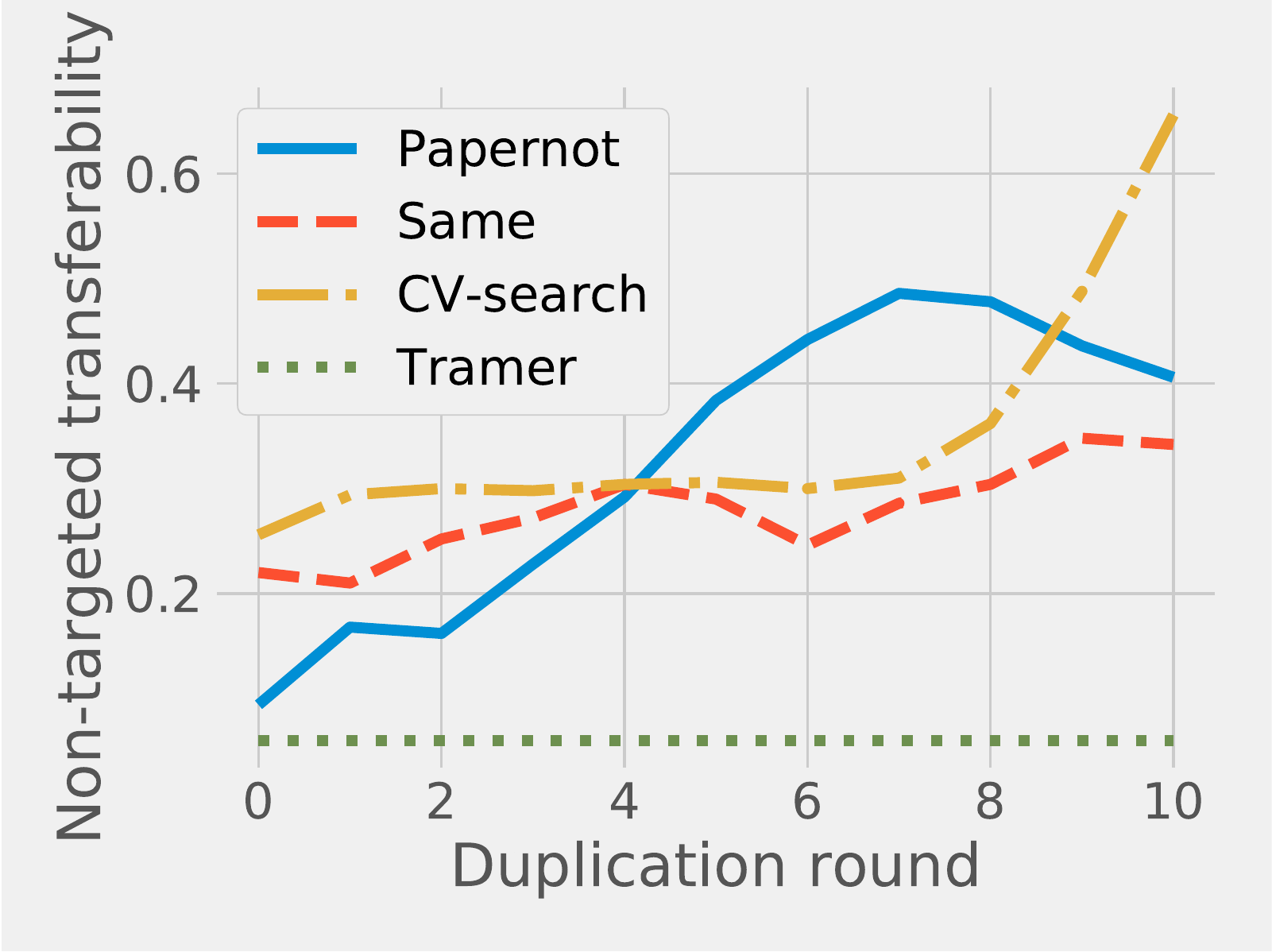}	
	\caption{\nontargeted transferability.}
	\end{subfigure}	
	\begin{subfigure}{0.23\textwidth}
	\centering
	\includegraphics[width=1.\columnwidth]{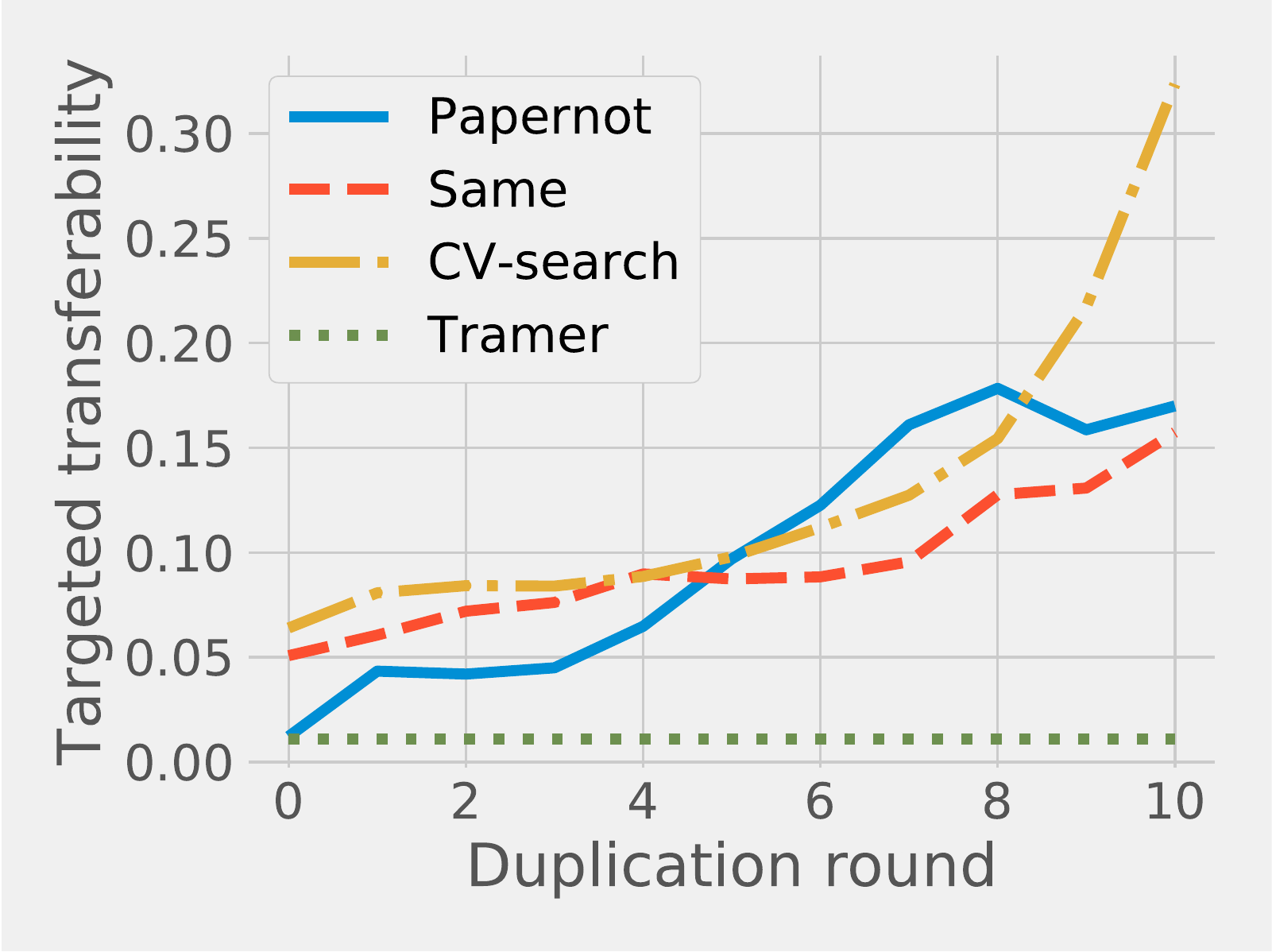}	
	\caption{\targeted transferability.}
	\end{subfigure}
	\caption{Model extraction performance vs. duplication rounds for four attack setups (Sect.~\ref{sec:pre-eval}) on MNIST. Mean results over 10 independent attacker sets. Transferability is significantly improved by using \cvsearch strategy both without synthetic data augmentation (duplication round 0), and after using 102,300 synthetic samples (duplication round 10).}
	  \label{fig:pre-eval}
\end{figure*}

We use the model architectures depicted in Tab.~\ref{table:architectures} for training our target models. At a high level, we separate three disjoint sets of data for our experiments:
\emph{test} set, \emph{target model training} set, and \emph{pre-attacker} set.
We call the set of initial seed samples in the model extraction process (Algorithm~\ref{alg:extraction}, row 6) the \emph{attacker} set, and it is a subset of \emph{pre-attacker} set. We vary its size systematically to understand the dependence of model extraction performance metrics on initial seed samples (Sect.~\ref{sec:seeds} for details). 

In MNIST we train 10 target models. The \emph{target model training} sets and \emph{pre-attacker} set are obtained by first separating the ``training'' set of MNIST with stratified 10-fold cross-validation, giving approximately a 6,000:54,000 split. The larger of these sets is used for \emph{target model training}. 
In GTSRB, we separated 36,629 images for \emph{target model training} and 2,580 for \emph{pre-attacker} set. GTSRB consists of up to 30 non-iid sequential samples of same physical objects photographed at different distances and angles. 
We ensured that same physical objects were only present in one of the datasets. We reserve the 12,630 test images for \emph{test} set. 
We trained all target models for 100 epochs using Adam~\cite{goodfellow2016deep} with learning rate 0.001. 
The learning rate was halved when the cost plateaud. 
The models reached on average 98\% accuracy on MNIST test set, and 95\% on GTSRB test set. 

\paragraph{Technicalities}

We measure the reproduction of predictive behavior with the \emph{agreement} metrics. It represents the accuracy of the substitute model predictions when taking the target model prediction as ground truth, i.e., a count of $\hat{F'}(x) = \hat{F}(x)$ occurrences.
We compute \testagree for a relevant subset of the input space as a macro-averaged F-score using MNIST and GTSRB test sets. This metric faithfully reports the effectiveness of an attack even in the case that classes are imbalanced. 
We compute the random uniform agreement \RUagree as an accuracy score on 4,000 samples chosen uniformly at random in the input space. 

We measure \emph{transferability} of adversarial examples over all seed samples in the attacker set. We measured both \targeted and \nontargeted transferability. We use the maximum perturbation $\epsilon = 64 / 255$ in our attack. For \targeted, we create 9 variants $x'$ of the initial sample $x$ with $\hat{F}(x) = c$, targeting 9 different classes $c' \neq c$. 

Adversarial examples can be crafted with a variety of maximum perturbations $\epsilon$. 
Larger values increase transferability, while impacting the visual perception of images to humans~\cite{sharif2018suitability}. 
The end-goal of our paper is not to discuss how these choices impact human perception. 
Creation of transferable adversarial examples simply 
serves as a way to evaluate the success of model extraction attacks. For this reason,
we choose the middle-range value of $\epsilon = 64/255$ throughout our paper. This value was the middlemost value evaluated in Papernot et al.~\cite{Papernot:2017:PBA}.

Our settings differ from prior works as follows: Tramer et al.~\cite{tramer:2016:stealing} did not evaluate 
model extraction attacks on DNNs; their largest neural network had $2,225$ parameters, while
our smallest network (MNIST) has $486,011$ parameters. 
The datasets they evaluated were smaller than the ones we use; our datasets are the same as in Papernot et al~\cite{Papernot:2017:PBA}.
Papernot et al. evaluated their attack up until $6,400$ queries, while we increase the total number of queries to $102,400$. \testagree in both earlier works is estimated with accuracy, while we use macro-averaged F-score to
faithfully report agreement for the underrepresented classes in the datasets, which is important in GTSRB. Finally, for GTSRB we ensured that the test set contained different physical images compared to the attacker set, whereas Papernot et al. did not.

\begin{figure*}
    \centering
    \begin{subfigure}[b]{0.23\textwidth}
        \includegraphics[width=\textwidth]{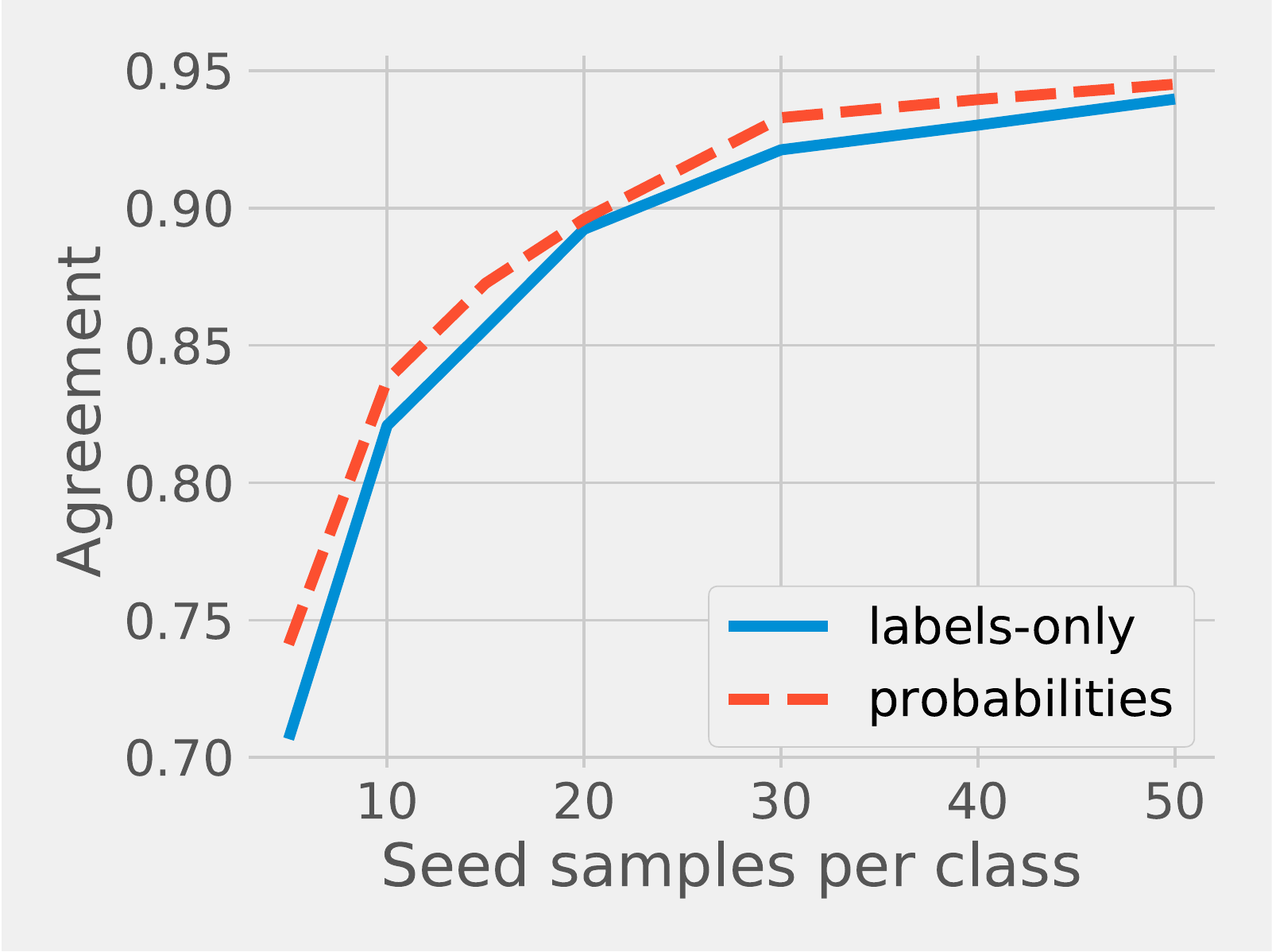}        
        \caption{MNIST \testagree.}
        \label{fig:mnist_a}
    \end{subfigure}
    ~ 
    \begin{subfigure}[b]{0.23\textwidth}
        \includegraphics[width=\textwidth]{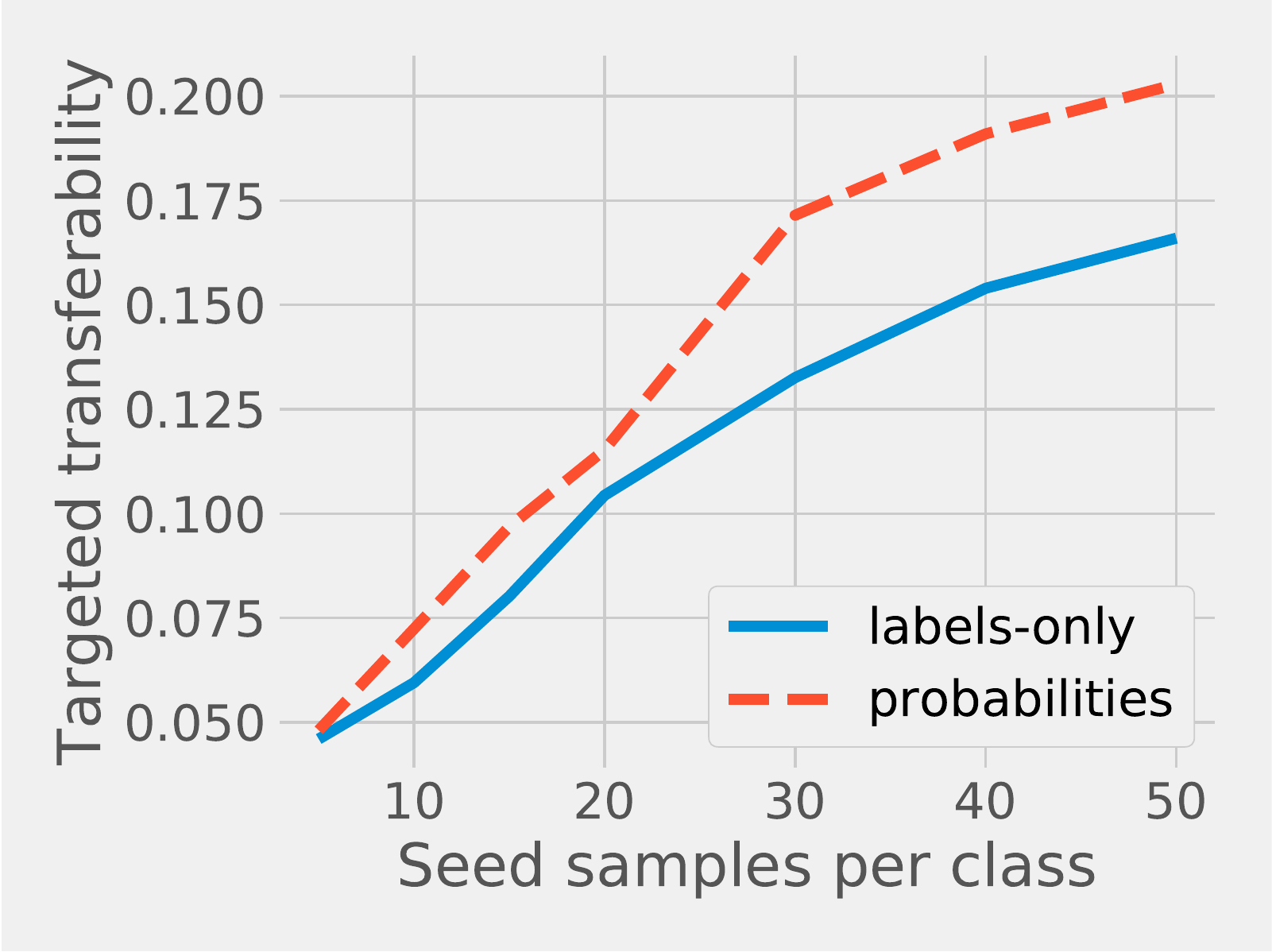}        
        \caption{MNIST \targeted.}
        \label{fig:mnist_t}
    \end{subfigure}
    ~ 
      \hspace{0.3cm}
    \begin{subfigure}[b]{0.23\textwidth}
        \includegraphics[width=\textwidth]{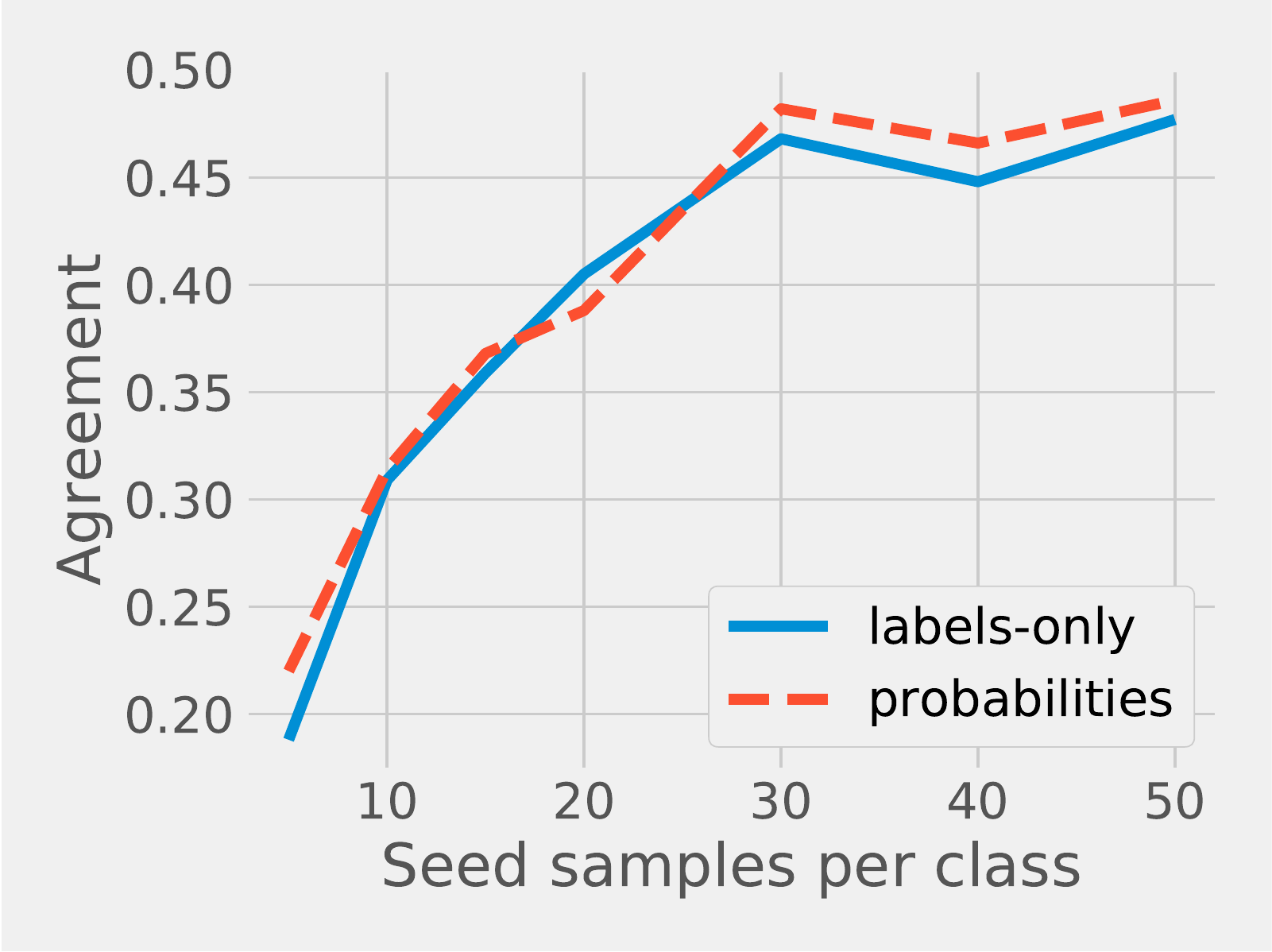}        
        \caption{GTSRB \testagree.}
        \label{fig:gtsrb_a}
    \end{subfigure}
    ~ 
    \begin{subfigure}[b]{0.23\textwidth}
        \includegraphics[width=\textwidth]{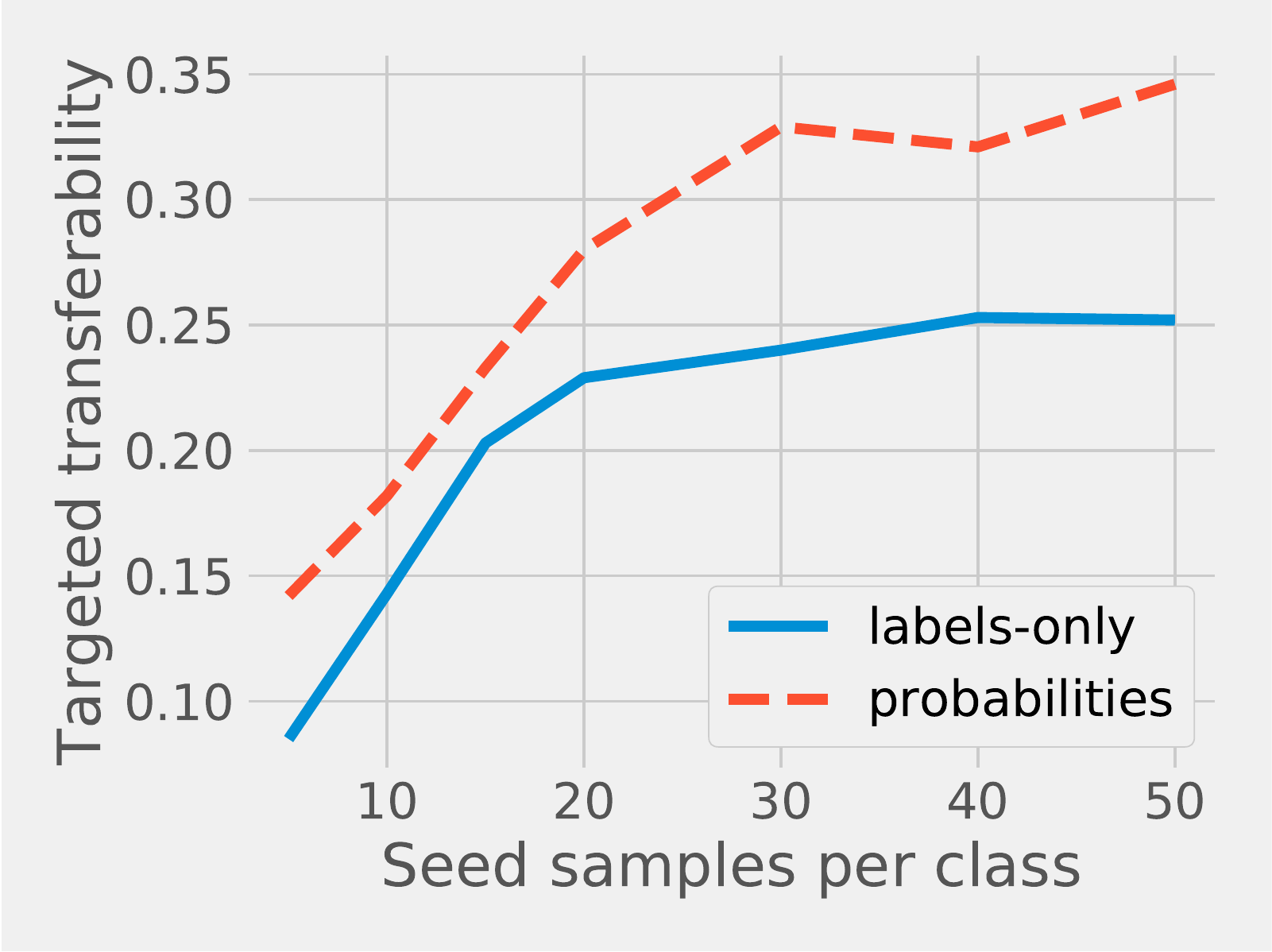}        
        \caption{GTSRB \targeted.}
        \label{fig:gtsrb_t}
    \end{subfigure}
        \caption{Effect of number of seed samples on model extraction performance on MNIST and GTSRB. 
        No synthetic samples are queried at this phase. The substitute model is trained with the \cvsearch strategy. }\label{fig:exp1}
\end{figure*}

\subsection{Evaluation of prior attacks and hyperparameters}
\label{sec:pre-eval}

We first evaluate \papernot and \tramer (Sect.~\ref{sec:prior}).
For brevity, we only report results on MNIST although the results for GTSRB showed similar results.
We report results for \papernot and \tramer on GTSRB further up in Sect.~\ref{sec:eval-end}. 

For \papernot, we use 10 natural seed samples per class (100 samples), and end after 10 duplication rounds ($2^{10}\cdot 100 = 102,400$ samples in total). 
\tramer initially queries random samples, followed by a line-search to find synthetic samples lying between existing ones, for a total of $102,400$ synthetic queries. 
We report the stronger \tramer, which uses probability outputs. 
Notably, \tramer uses 100\% synthetic samples, whereas \papernot use 0.1\% natural samples and 99.9\% synthetic samples. 
In addition to these, we test two new variants of \papernot with new hyperparameters (Sect.~\ref{sec:hyperparameters}). 


At this state, we only vary the hyperparameter setup. 
We show the evolution of agreement and transferability as the number of duplication rounds increases in Fig.~\ref{fig:pre-eval}. 
\tramer only uses one duplication round, so we show it as a straight line for clarity.
Transferability is computed using MI-FGSM with step size $\epsilon = 64/255$. 
There are several interesting observations: 1) \tramer produces the best \RUagree, but the performance translates neither to good \testagree, nor to good transferability. 
2) both initial agreement and transferability are highest for
models trained with \cvsearch, 
even higher than on \same. 
3) \testagree on MNIST is ultimately dominated by the high volume of synthetic samples in the attacker set: all training setups converge to the same \testagree. 
4) \RUagree is initially random (Fig.~\ref{fig:pre-eval}) -- the accuracy of these is approximately $\frac{1}{C} = 10\%$, where $C$ is the number of classes. The
more synthetic queries are sampled, the higher the \RUagree rises. 
However, the increase stagnates at the sixth duplication round (at 6400 samples), and
does not rise further for any method. We suspect that this is due to the limitations of FGSM (recall discussion regarding Fig.~\ref{fig:synthetics_schema} in Sect.~\ref{sec:synthetic_generation}).
5) \cvsearch is the only method where transferability starts improving exponentially after the seventh duplication epoch.
We verified that this effect is due to \emph{dropout training}. Other models may
not improve due to overfitting model parameters on the substantial \emph{attacker}
 set data, and dropout may help in avoiding this phenomena.  
6) \papernot is the fastest attack among these: querying and training (without network latency) took on average 4.5 minutes, while it took 26 min and 18 min respectively for the \cvsearch and \same attacks.

We report detailed transferability for \papernot in Tab.~\ref{table:setup1} after the last duplication round,
calculated with different adversarial example crafting algorithms (Sect.~\ref{sec:adversarial_example}), including the FGSM evaluated in~\cite{Papernot:2017:PBA}.
The iterative algorithms are run with 11 steps each. 
We also show a \emph{random} perturbation of the same size for comparison. 
To be brief, we only report transferability for \papernot, 
although the results for the different attacks showed the same pattern.
The actual numbers for transferability differs from~\cite{Papernot:2017:PBA},
as the target model architecture they attacked is not disclosed. 

\begin{table}[h]
  \begin{center}
    \begin{tabular}{c c c c} \hline
    \textbf{Evasion method} & \textbf{\nontargeted} (\%) & \textbf{\targeted} (\%) & $\mathbf{L_2}$ \textbf{norm} \\ \hline
    random & $7.5 \pm 2.7$ & $1.0 \pm 0.4$ & $10.3 \pm 0.01$ \\

    FGSM & $24.4 \pm 5.9$  &  $8.3  \pm 2.0$ & $9.2  \pm 0.04$ \\
    I-FGSM & $57.4 \pm 7.1$ &  $12.2 \pm 2.3$ & $9.3  \pm 0.04$ \\
    MI-FGSM & $40.6 \pm 8.4$ & $17.0 \pm 4.3$ & $8.5  \pm 0.05$ \\
    \hline
    \end{tabular}
    \caption{MNIST. Transferability of \papernot after 10 duplication rounds, evaluated with different evasion attacks. We use $L_{\infty}$ bound $\epsilon = 64 / 255$ for transferability calculation. $L_2$ norm is shown for \nontargeted, for all generated samples. Mean and standard deviations shown.}
    \label{table:setup1}
  \end{center}
\end{table}  

We find that although \papernot uses FGSM to craft synthetic samples, transferability
is better when crafting adversarial examples using the iterative variants of the algorithm:
 I-FGSM and MI-FGSM.
We can see that iterative variants of the attack are stronger in our scenario.  
After synthetic queries had been queries, we observed that \nontargeted is $2 \times$, 
and \targeted up to $3 \times$ higher for iterative variants of the attack. 
For this reason, transferability will only be evaluated using iterative variants
of the attack in the rest of this paper.



\subsection{Impact of seed samples}
\label{sec:seeds}

Next, we explore the connection between the number of seed samples and model extraction efficiency,
to understand the impact that these have on overall attack efficiency. 
We use the \cvsearch training strategy, which we demonstrated worked the best, and do not query synthetic data at this stage.
We also investigate whether more detailed information from the target models -- 
the full list of classification \emph{probabilities} rather than the top-1 \emph{labels-only}
-- can aid in the model extraction. 
These may be considered the highest and respectively lowest levels of granularity that any prediction API may provide. 


Figure~\ref{fig:exp1} shows \testagree, and \targeted for MNIST and GTSRB, as the number of training samples (natural seed samples) increases. Transferability is calculated with MI-FGSM. Overall \testagree trends are similar on both datasets, but \testagree is smaller on GTSRB than on MNIST. We believe this is due to higher dissimilarity between samples in attacker set, and target model training set. 

\testagree does not improve significantly with probabilities; we observe increases between 0 and 3 percentage points (pp). This is in contrast with findings on shallow architectures in~\cite{tramer:2016:stealing}, where probabilities increased model extraction efficiency significantly. 
Increasing the number of training samples to 10-fold (from 5 samples per class to 50 samples per class) increases \testagree by 23 pp on MNIST and 29 pp on GTSRB, reaching 93\% and 47\% \testagree respectively. 

\targeted improves with probabilities. 
The effect is more pronounced, when the adversary has access to more seed samples. 
For MNIST, 
\targeted starts at 5\% when there are only 5 seed samples per class, and reaches 16\% with 50 seed samples per class with labels-only, and 20\% with probabilities. 
Perhaps surprisingly, \targeted on GTSRB is higher than on MNIST. 
\targeted stagnates in GTSRB after 20 -- 30 seed samples per class. We believe that the stagnation occurs due to correlated samples in the seed sample set, 
due to the structure of GTSRB dataset. 
When 50 samples per class have been queried, \targeted reaches 25\% with labels-only, and 35\% with probabilities. 

\nontargeted (not shown) is behaving similarly on MNIST: 
training with either labels-only or probabilities
yields 20\% transferability with 5 seed samples per class. 
When 50 samples per class have been queried, \nontargeted reaches 48\% with labels-only,
and 65\% with probabilities.
\nontargeted on GTSRB already starts at 91\% with labels-only, and can reach 98\% with 50 seed samples per class. \nontargeted is already very high, and probabilities yields at most 4 pp improvements over labels-only.

Having demonstrated the overall trend of increasing seed sample numbers and impact of probabilities, we will investigate the settings with 10 seed samples per class in more detail in the following sections.





\subsection{Synthetic sample generation}
\label{sec:synthetic-eval}

We explore the impact that different synthetic sample crafting settings have on model extraction efficiency.
As in the previous tests, we use \cvsearch, trained with 10 seed samples per class. 
At this phase, we only evaluate the scenario where the adversary has access to labels. 
We run model extraction attacks against MNIST and GTSRB using several synthetic sample
crafting techniques (Section~\ref{sec:synthetic_generation}). 
\trnd and \colorr techniques are used with expansion factor $k=4$. 
Step size $\lambda$ is set to $25.5/255$, and adversarial examples are crafted with I-FGSM. 

\begin{table}[h]
  \begin{center}
    \begin{tabular}{c c c c} \hline
    \textbf{Synthetic crafter} & \textbf{\testagree} & \textbf{\targeted} & \textbf{\nontargeted} \\ \hline
 
         N FGSM &           0.960 &           0.283 &           0.770 \\ \hline
        N I-FGSM &          -0.001 &          -0.006 &          -0.086 \\
   T-RND FGSM &           \textbf{+0.008} &           \textbf{+0.046} &           0.008 \\
  T-RND I-FGSM &           +0.007 &           0.043 &           \textbf{+0.056} \\
   COLOR &          -0.071 &          -0.219 &          -0.500 \\
    \hline
    \end{tabular}
    \caption{Impact of synthetic sample crafting strategy on model extraction performance. MNIST, 102,400 queries.}
    \label{table:exp2_m}
  \end{center}
\end{table}

We first discuss the results for model extraction attacks on MNIST, shown in Tab.~\ref{table:exp2_m}. 
Non-targeted FGSM is kept as the baseline, and other methods are compared against this setup.
Unsurprisingly, \targeted is most increased by using targeted synthetic sample
crafting methods (T-RND), on average by 4.5 pp. 
\nontargeted also increases by 5.6 pp using T-RND I-FGSM. 
COLOR decreases \testagree and transferability over the baseline.
\testagree results are already quite high for the baseline method, but targeted methods
provide nearly one pp improvement over the baseline.


\begin{table}[h]
  \begin{center}
    \begin{tabular}{c c c c} \hline
    \textbf{Synthetic crafter} & \textbf{\testagree} & \textbf{\targeted} & \textbf{\nontargeted} \\ \hline
         N FGSM &           0.396 &           0.593 &           1.000 \\ \hline 
        N I-FGSM &           +0.056 &          -0.075 &           0.000 \\
   T-RND FGSM &          -0.135 &           +0.002 &           0.000 \\
  T-RND I-FGSM &           +0.112 &           \textbf{+0.170} &           0.000 \\
   COLOR &           \textbf{+0.243} &          -0.498 &          -0.016 \\
    \hline
    \end{tabular}
    \caption{
    Impact of synthetic sample crafting strategy on model extraction performance. GTSRB, 110,800 queries. }
    \label{table:exp2_g}
  \end{center}
\end{table}  

Results for the attack on GTSRB are shown in Tab.~\ref{table:exp2_g}. 
\nontargeted is already 100\% on the baseline, and none of the Jacobian-based Synthetic Sample Generation methods
decrease it. 
We observe that creating synthetic samples using targeted methods increases
\testagree and \targeted over baseline, with T-RND I-FGSM contributing the largest
 increase in \targeted. 
The largest impact of \testagree comes from the domain-specific method COLOR. 
We hypothesize that this is due to images occasionally having very large differences in contrast in attacker set samples, and randomly changing colors may help in bridging the gap between attacker set images and target model training set. 

We further found that large values for $\lambda$ were beneficial in MNIST,
both in terms of \testagree and transferability. On GTSRB, larger $\lambda$
improved transferability as well, but decreased \testagree.
In the next section, we evaluate these two goals separately: doing model stealing
with large $\lambda$ for transferability, and smaller $\lambda$ for \testagree. 


\subsection{Comparative evaluation to prior work}
\label{sec:eval-end}

We summarize the performance of existing model extraction techniques, 
and our techniques in Tab.~\ref{table:summary}. We show \testagree and
Transferability for our two datasets in two scenarios: using no synthetic queries
and using approximately 100,000 synthetic queries. 
We set the number of natural seed samples to 10 per class. 
Transferability is evaluated with I-FGSM. 


For MNIST, we find that our \cvsearch technique yields comparable \targeted as \papernot even before synthetic samples are queried. 
With a budget of 102,400 queries, 
\papernot reaches average \testagree 95.1\%, while
T-RND I-FGSM with step size $\lambda=64/255$ reaches 97.9\%, while \targeted and \nontargeted(not shown), increase to 39.9\% and 87.7\% compared to 10.6\% and 56.2\% respectively in \papernot. 
Our techniques improve \targeted and \nontargeted on MNIST by +29.3 pp and 
+31.5 pp.

\begin{table}[t]
  \begin{center}
    \begin{tabular}{c c c|c c} \hline
    & \multicolumn{4}{c}{MNIST}  \\
    & \multicolumn{2}{c}{No synthetic queries} & \multicolumn{2}{c}{102,400 total queries} \\
   \textbf{Strategy} & \multicolumn{1}{c}{\textsl{\textbf{Test-agree.}} } & \multicolumn{1}{c}{\textbf{\targeted}} & \multicolumn{1}{c}{\textsl{\textbf{Test-agree.}} } & \multicolumn{1}{c}{\textbf{\targeted}}\\ \hline
         \tramer &     -  & - & 6.3\% & 1.1\% \\ 
         \papernot &   40.0\% & 1.2\% & 95.1\% & 10.6\% \\ 
         Our T-RND-64 & \textbf{82.9\%} & \textbf{6.5\%} & \textbf{97.9\%} & \textbf{39.3\%}  \\          
    \hline
    & \multicolumn{4}{c}{GTSRB}  \\
    & \multicolumn{2}{c}{No synthetic queries} & \multicolumn{2}{c}{110,880 total queries} \\ \hline
         \tramer &     -  & - & 0.2\% & 2.1\% \\ 
         \papernot & 4.8\% & 2.4\% & 16.9\% & 41.1\% \\ 
         Our T-RND-64 & \textbf{32.0\%} & \textbf{16.9\%} & 47.6\% & \textbf{84.8\%} \\          
         Our COLOR-25 & \textbf{32.0\%} & \textbf{16.9\%} & \textbf{62.5\%} & 27.5\% \\          
    \hline
    \end{tabular}
    \caption{
    Comparative evaluation of model extraction attacks on our two datasets.
    Our techniques achieve significantly improved performance on both \testagree
    and \targeted. 
}
    \label{table:summary}
  \end{center}
\end{table}

We see that the \cvsearch technique we employ is crucial for \testagree on GTSRB: 
\papernot with 110,800 queries does not reach the same \testagree as our techniques reaches without synthetic samples. 
\emph{T-RND} with step size $\lambda = 64/255$ further increases \testagree to 47.6\%, while \targeted increases to 84.8\%, compared to 41.1\% in \papernot. \nontargeted (not shown) is 100.0\% in both cases. 
\emph{COLOR} increases \testagree to 62.5\%, while decreasing \targeted to 14.9\%, highlighting that achieving high \testagree may be entirely complimentary to achieving transferability in model extraction attacks. 
\tramer performs poorly on both datasets. This is because a large part of the random space 
belongs to only one class; this information is unhelpful towards building a good substitute model. 
Our techniques improve \testagree and \targeted on GTSRB by +46 pp 
and +44 pp, respectively. 

\subsection{Architecture mismatch between target and substitute models}
\label{sec:complexity}

Having seen the effect that various training strategies and synthetic sample crafting techniques have,
we may ask what happens if the attacker does not know the target model architecture and instead, uses a simpler or more complex substitute model architecture
than the target architecture.

\begin{figure}
    \centering
        \begin{subfigure}[b]{0.5\textwidth}
        \includegraphics[width=\textwidth]{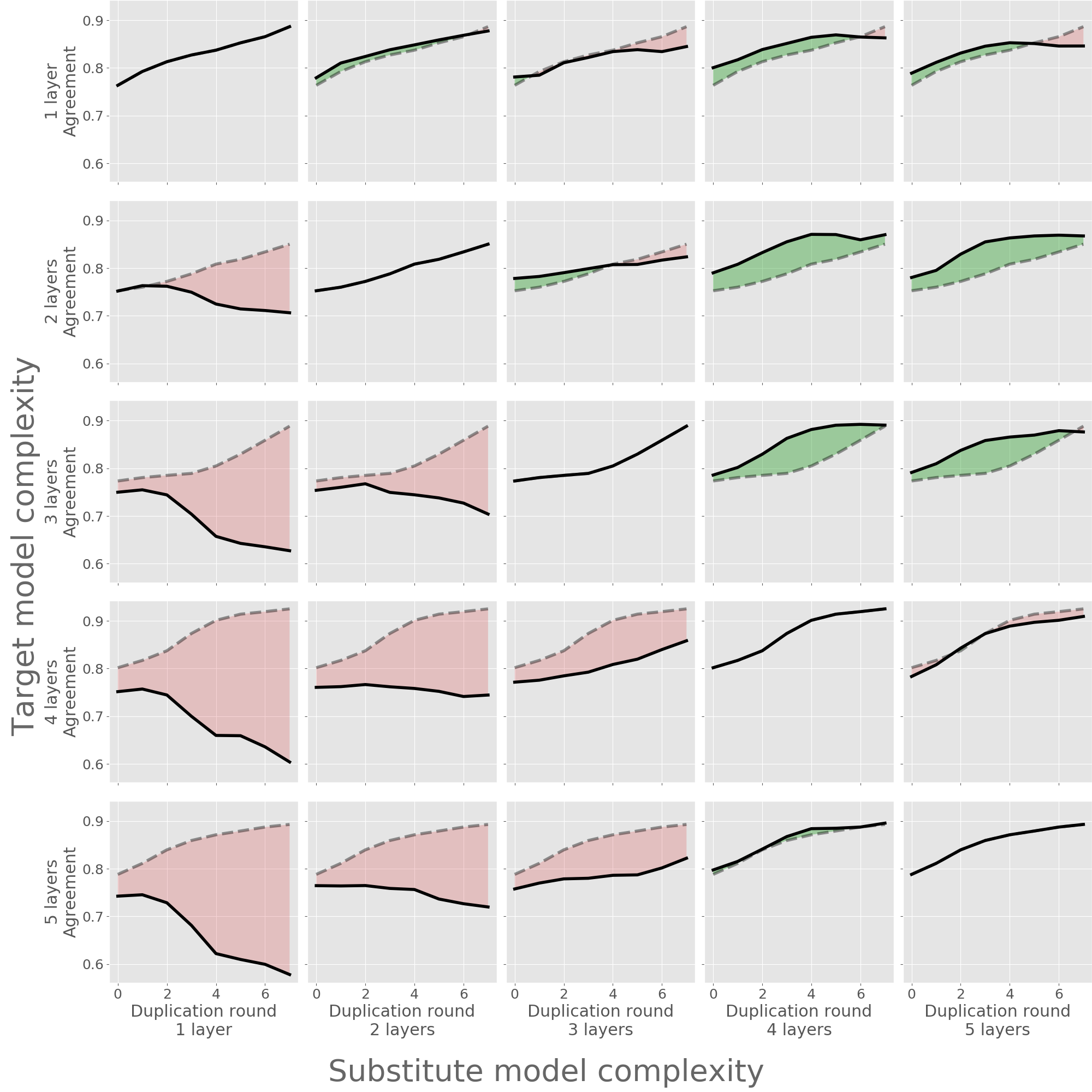}
        \end{subfigure}
        \caption{
        Effect of architecture mismatch on \testagree (cf. Tab.~\ref{table:layers}). 
        Columns represent increasing complexity of the substitute model (left to right).
        Rows represent increasing complexity of the target model (top to bottom). 
        Substitute models with lower complexity than the target model (lower triangle, pale red) have significantly lower \testagree, compared to the baseline of matching architectures (diagonal). 
        }\label{fig:agreement_model_complexity}
\end{figure}

We trained target models of 5 different complexities on MNIST, 
using architectures detailed in Tab.~\ref{table:layers}. 
The architectures are chosen for simplicity of evaluating increased (nonlinear) complexity. 
\techreport{Architectures ``3 layers'', ``4 layers'' and ``5 layers'' contain convolutional layers. }{}
We trained each substitute model with the \same method, and used the non-targeted FGSM for crafting synthetic samples with 7 duplication rounds.
We ran the attack ten times for each of the different complexities.   
We show our average results for \testagree in Fig.~\ref{fig:agreement_model_complexity}. 
Columns denote increasing substitute model complexity as we move rightwards, 
and rows denote increasing target model complexity as we move downwards. 
The baseline is the diagonal, where model complexities match. We hilight row-wise positive change over the baseline with solid green, and negative change with pale red. 

\begin{table}[h]
  \begin{center}
    \begin{tabular}{ c c c c c } \hline
    \multicolumn{5}{c}{\textbf{Architecture}} \\
    1 layer & 2 layers & 3 layers & 4 layers & 5 layers \\ \hline

      & & & & conv2-32 \\ 
      & & & & maxpool2 \\\hline
      & & & conv2-32 & conv2-64 \\ 
      & & & maxpool2 & maxpool2 \\  \hline
      & &conv2-32 &conv2-64 & conv2-128\\ 
      & &maxpool2 &maxpool2 & maxpool2 \\\hline
      &FC-200 &FC-200 &FC-200 & FC-200\\ \hline
    FC-10 & FC-10 & FC-10 & FC-10 & FC-10 \\ \hline  \\
    \multicolumn{5}{c}{\textbf{Parameters}} \\ \hline
    7,851 & 159,011 & 1,090,171 & 486,011 & 488,571 \\ \hline

    \end{tabular}
    \caption{Different model architectures for analysing architecture mismatch in target and substitute models. ReLU activations are used between blocks of layers. The number of parameters in the networks in reported at the bottom.}
    \label{table:layers}
  \end{center}
\end{table}

We see a clear pattern: To increase \testagree, matching or having higher model complexity than the target model is almost always beneficial for the adversary. Similarly, using a lower complexity is detrimental to the attacker, 
and can cause a breakdown of the attack, where \testagree drops lower than initially. 
This phenomena may be explained via statistical learning theory, which provides 
an impossibility result of perfectly reproducing a high-complexity classifier
with a classifier of too low complexity, i.e. too low Vapnik-Chervonenkis (VC) dimension~\cite{murphy2012machine}. 


\nontargeted and \targeted transferability are also affected by
model mismatch, but in a different way. 
We show the results for \nontargeted in Tab.~\ref{table:layers_transferability},
similarly evaluated on duplication round 7 using I-FGSM with step size $\epsilon = 64/255$. 
We see that nearly all (19 out of 20) deviations from the target model architecture cause significant decrease in the ability to produce transferable adversarial examples.


   

\begin{table}
  \begin{center}
    \begin{tabular}{c c c c c c} \hline
&\multicolumn{5}{c}{\textbf{Baseline \nontargeted} (\%)} \\ \hline
& 99.4 & 64.6 & 78.4 & 36.3 & 15.5 \\ 
\hline \\
            & \multicolumn{5}{c}{\textbf{Relative Improvement} (\%)} \\       
 & 1 & 2 & 3 & 4 & 5  \\
       \hline       
1 layer~ &0.0 & -0.7 & -34.8 & -46.7 & -38.4 \\
2 layers &-70.0 & 0.0 & -73.5 & -75.5 & -61.0 \\
3 layers &-85.2 & -84.9 & 0.0 & -49.4 & -52.9 \\
4 layers &-72.2 & -58.1 & -24.8 & 0.0 & -0.3 \\
5 layers &-67.7 & -67.1 & -29.7 & +8.4 & 0.0 \\
  
       \hline    
    \end{tabular}
    \caption{Effect of architecture mismatch on improvement on \nontargeted
    transferability. Columns represent increasing complexity of the substitute model (left to right).
    Rows represent increasing complexity of the target model (top to bottom). Matching architectures (diagonal) improves transferability. }
    \label{table:layers_transferability}
  \end{center}

\end{table}


\subsection{Takeaways}

We conducted systematic, empirical tests to understand model extraction attacks on DNNs in the previous sections. We present our main observations as follows:

\noindent\textbf{Hyperparameters}: It is not necessary to use the same learning rate and number of training epochs as the target model was trained with. Doing \cvsearch can yield similar or better results for both agreement and transferability. 

\noindent\textbf{Seed samples}: \techreport{Natural seed samples are necessary to extract a substitute model that has good \testagree: the more natural samples the adversary has the higher the \testagree.}{Natural seed samples are necessary to extract a substitute model that reaches high \testagree.}


\noindent\textbf{Synthetic sample generation}: A relevant synthetic sample generation method improves transferability of adversarial examples significantly. 
Synthetic samples also significantly improve agreement, while remaining less efficient than using natural samples.
Exploring several directions (T-RND) yields better agreement and transferability. 

\noindent\textbf{Training strategy}: The use of probabilities rather than labels-only improves transferability for any setup, but has nearly no effect on agreement. 

\noindent\textbf{Mismatch between target model and substitute model}: Using a higher or similar complexity substitute model as the target model architecture yields high predictive performance. Matching the architectures yields higher transferability. 

\noindent\textbf{Generalizability}: Our dataset choices facilitated comparisons with existing methods, where they had poor \textsl{Test-agreement}: \cite{Papernot:2017:PBA} on GTSRB and \cite{tramer:2016:stealing} on both. 
In these adversary models, adversary is not assumed to have access to \emph{pre-trained models}: 
both the target and the substitute model  
DNNs are trained \emph{from scratch}. 
Since the time of this writing, stealing DNNs for more complicated datasets like CIFAR-10\footnote{cf.~\url{http://karpathy.github.io/2011/04/27/manually-classifying-cifar10/}}, 
have been done by assuming \emph{both} the target model and attacker models are 
fine-tuned from pre-trained ImageNet classifiers~\cite{DBLP:journals/corr/abs-1812-02766, pengcheng2018query}. These attacks benefit from correlations between different~\cite{pengcheng2018query} or same~\cite{DBLP:journals/corr/abs-1812-02766} pre-trained models. 
In contrast, our paper analyzes attacks where no such correlation is present. 


\section{Detecting Model Extraction}
\label{sec:detection}
We present \ourname (\underline{Pr}otecting \underline{a}gainst \underline{D}NN Model Stealing \underline{A}ttacks), a generic approach to detect model extraction attacks. Unlike prior work on adversarial machine learning defenses, e.g., for detecting adversarial examples~\cite{grosse2017statistical,Meng2017magnet}, our goal is not to decide whether individual queries are malicious but rather detect attacks that span several queries. Thus, we do not rely on modeling what queries (benign or otherwise) look like but rather on how successive queries relate to each other.
\ourname is generic in that it makes no assumptions about the model or its training data.


\subsection{Detection approach}
\label{sec:detection-approach}

We start by observing that (1) model extraction requires making several queries to the target model and (2) queried samples are specifically generated and/or selected to extract maximal information.
Samples submitted by an adversary are expected to have a characteristic distribution that differs from the distribution of samples submitted in benign queries.

The distance between two randomly selected points from a totally bounded space (e.g., a cube) almost fits a normal (Gaussian) distribution~\cite{philip2007probability,sors2004integral}. Inputs to a machine learning model are typically defined in a totally bounded input space, i.e., input features are defined over a certain range of values.
We expect benign queries from a given client to be distributed in a natural and consistent manner. Consequently, we expect the distance between queried samples to fit a (close to) normal distribution, as observed for random points.
On the other hand, adversarial queries made to extract a model combine natural and synthetic samples coming from different distributions. Moreover, the distance between successive synthetic queries is artificially controlled by the adversary to optimally probe the input space and extract maximal information~\cite{Papernot:2017:PBA,tramer:2016:stealing}.
Therefore, we expect the distance between adversarial queries to highly deviate from a normal distribution.

Thus, \ourname's detection method is based on detecting deviations from a normal distribution in the distance between samples queried by a given client.

\begin{figure}[ht]
	\begin{subfigure}{0.49\columnwidth}
	\centering
	\includegraphics[width=\columnwidth]{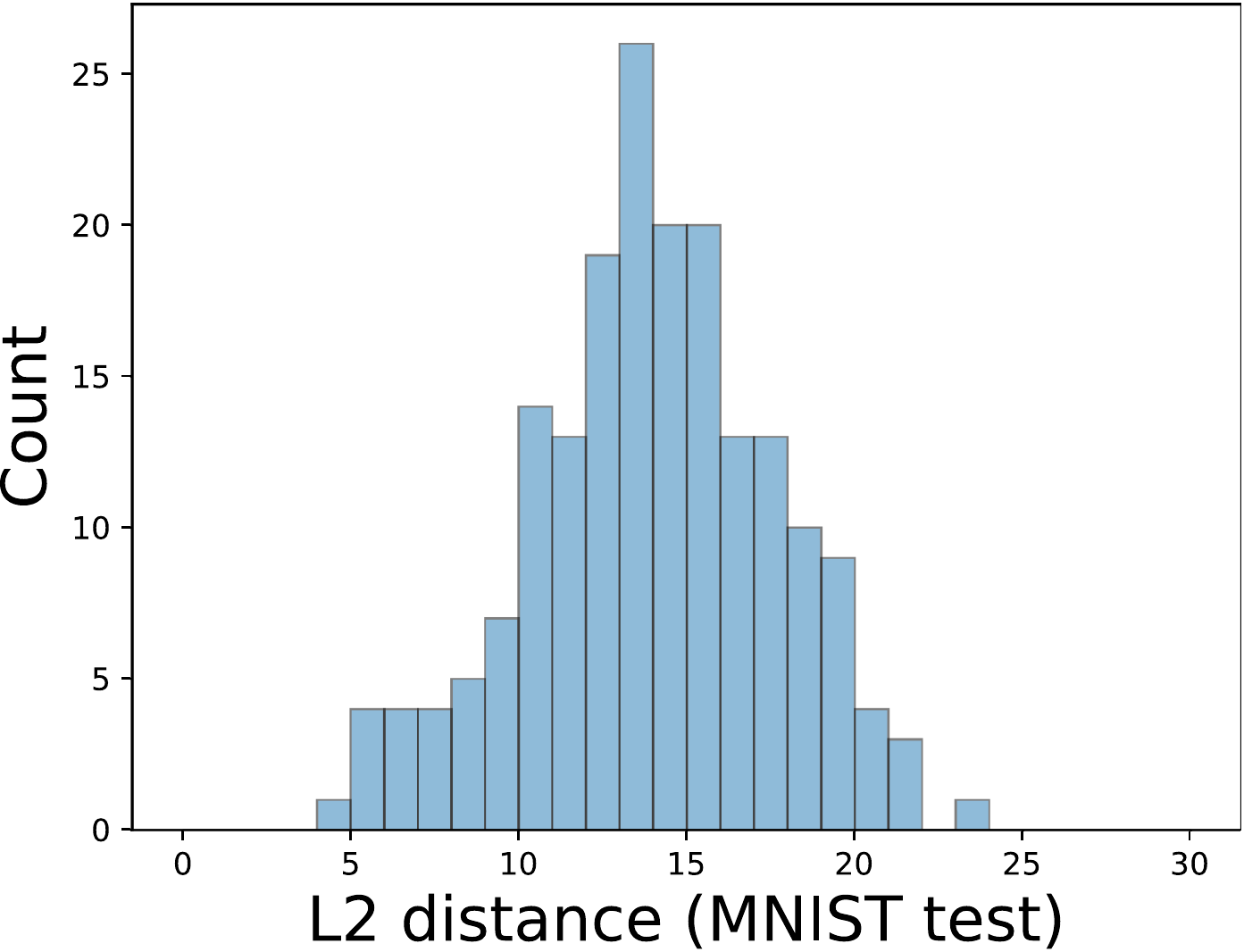}
		\captionof{figure}{Benign queries (MNIST)}
	\label{fig:mnist_test}
	\end{subfigure}
	\begin{subfigure}{0.49\columnwidth}
	\centering
	\includegraphics[width=\columnwidth]{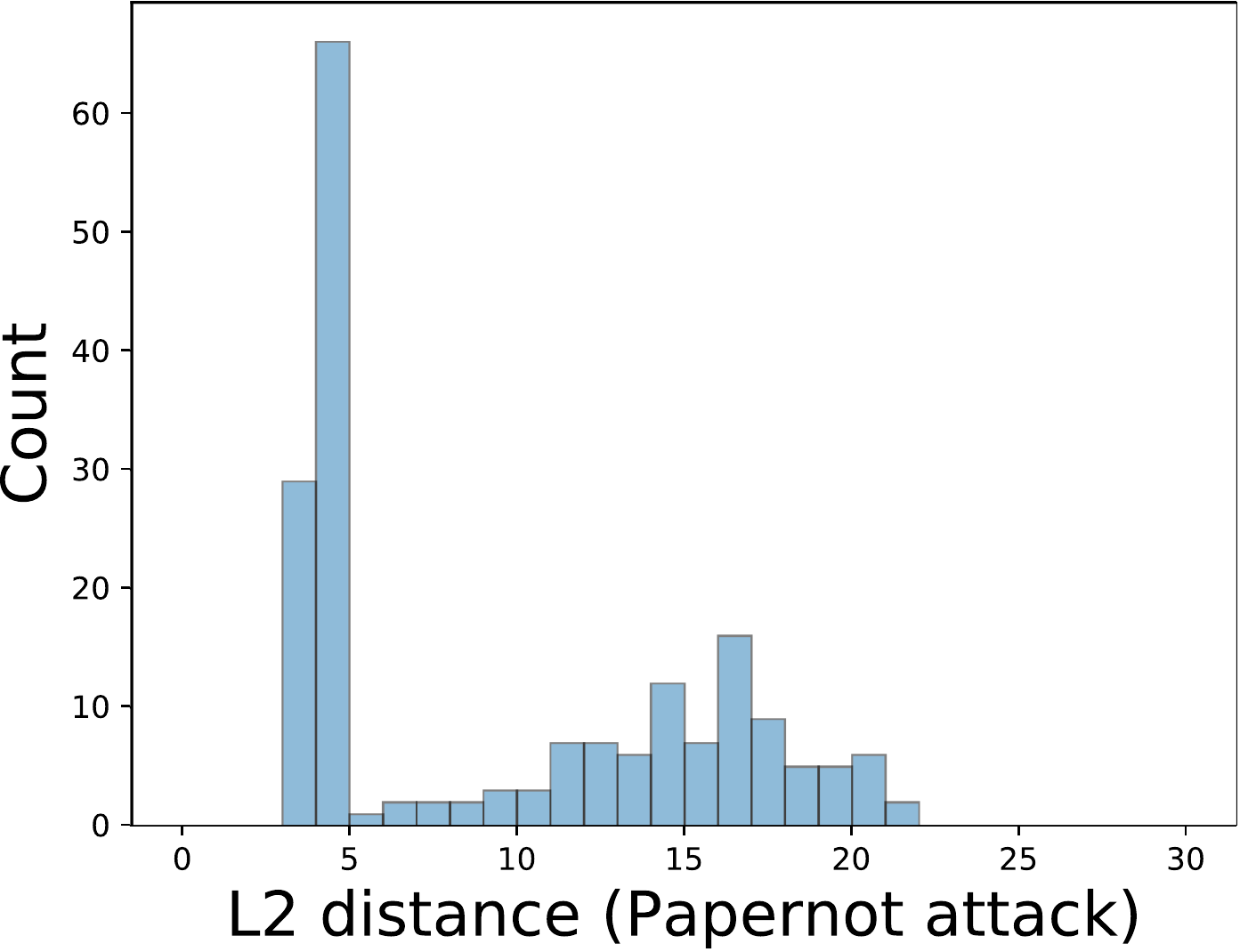}
		\captionof{figure}{\papernot attack (MNIST)}
	\label{fig:mnist_pap}
	\end{subfigure}
		\begin{subfigure}{0.49\columnwidth}
	\centering
	\includegraphics[width=\columnwidth]{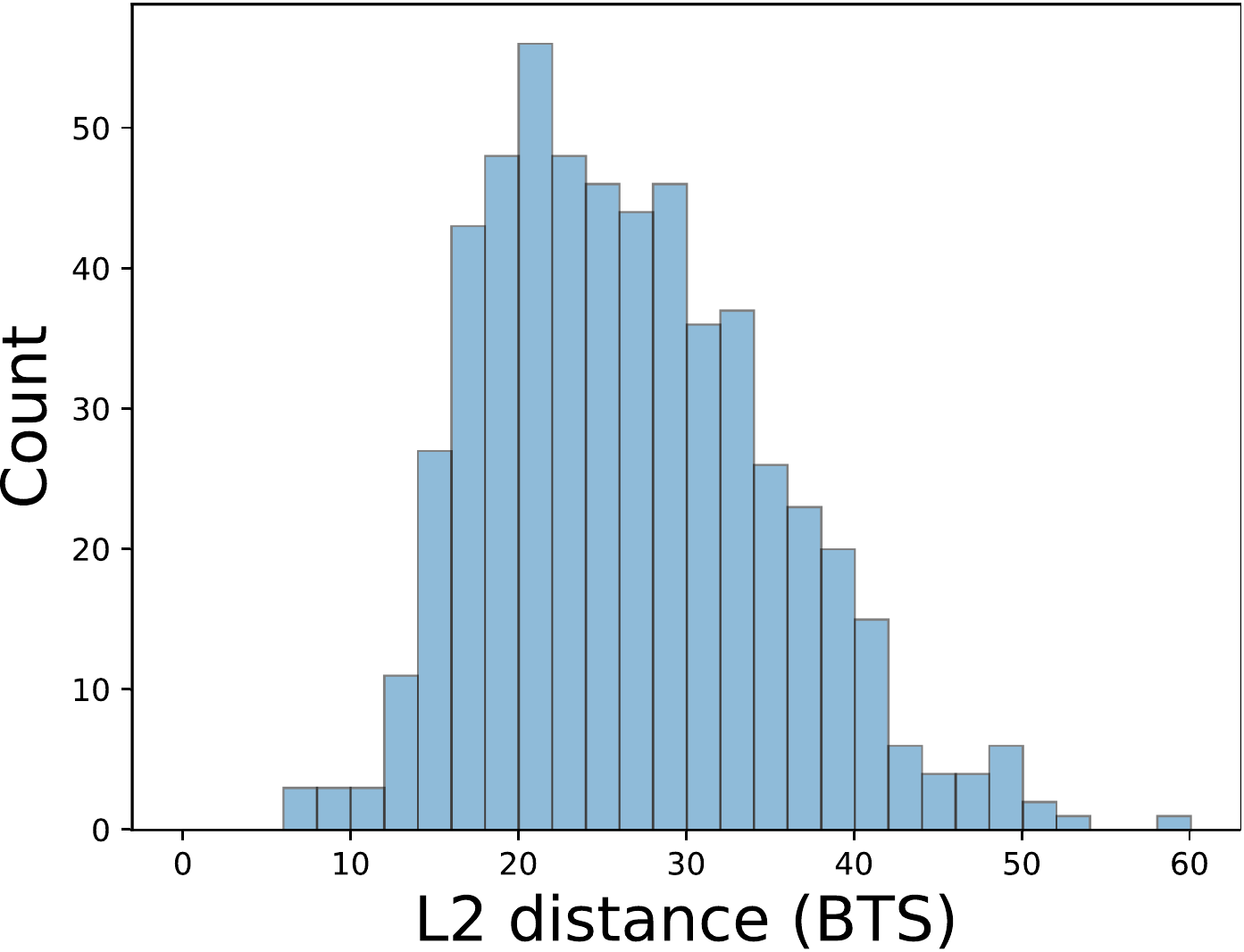}
		\captionof{figure}{Benign queries (GTSRB)}
	\label{fig:gtsrb_bts}
	\end{subfigure}
	\begin{subfigure}{0.49\columnwidth}
	\centering
	\includegraphics[width=\columnwidth]{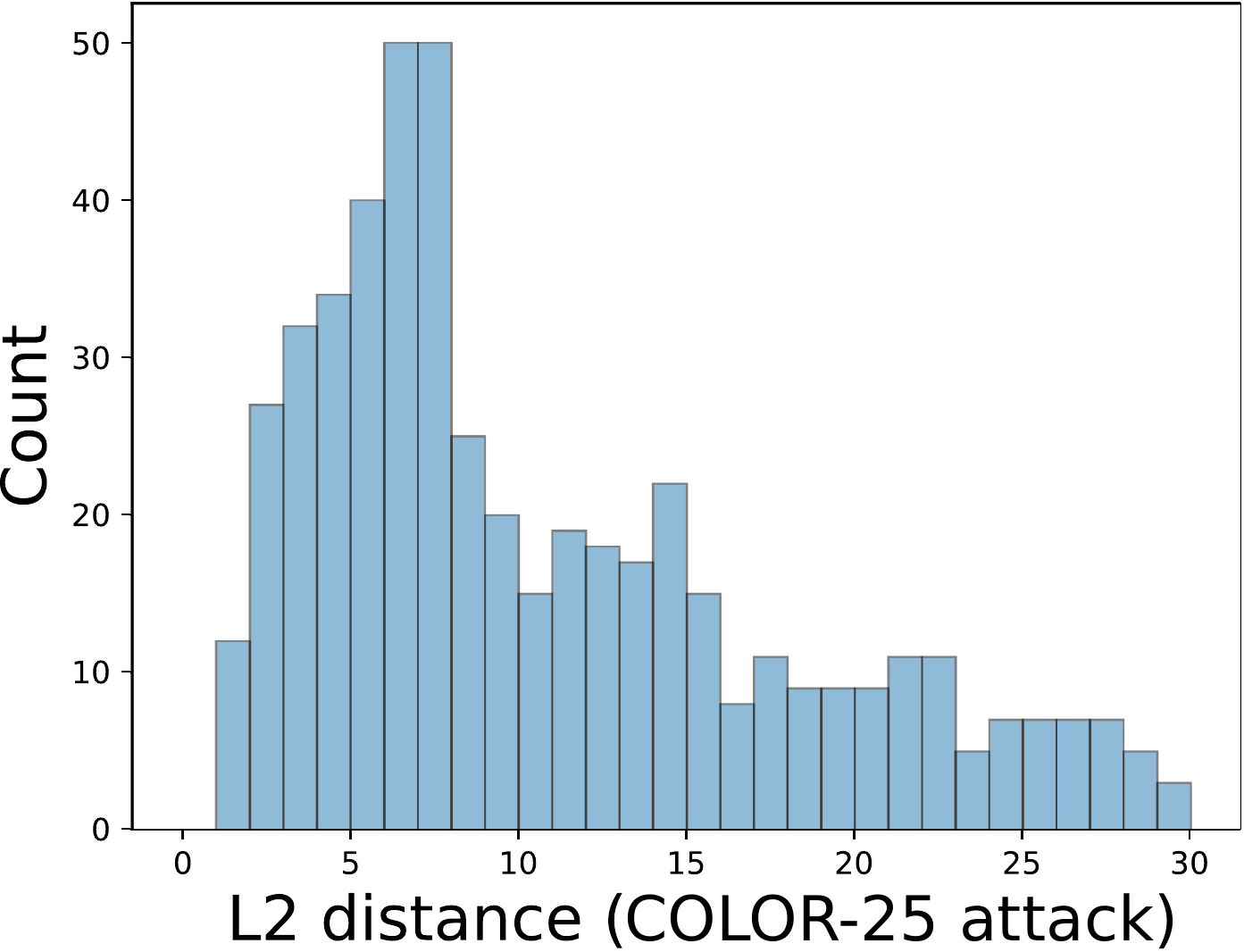}
		\captionof{figure}{\colorr attack (GTSRB)}
	\label{fig:gtsrb_trnd}
	\end{subfigure}
	\caption{\label{fig:query_distrib}Distribution of distances for benign queries and adversarial queries. \textbf{Top:} 200 queries against MNIST model, benign queries from MNIST test set (left) and \papernot attack (right). \textbf{Bottom:} 600 queries against GTSRB model, benign queries from BTS dataset (left) and \colorr attack (right). Benign queries have a distribution close to normal while adversarial queries do not.}

\end{figure}

\begin{algorithm}[ht]
\caption{\ourname's detection of model extraction}
\label{algo:detection}
\begin{algorithmic}[1]
	\State{
		Let $F$ denote the target model,
		$S$ a stream of samples queried by a given client,
		$D$ the set of minimum distances $d_{min}$ for samples in $S$,
		$G_c$ the growing set for class $c$,
		$D_{Gc}$ the set of minimum distances $d_{min}$ for samples in $G_c$,
		$T_c$ the threshold value for class $c$,
		$\delta$ the detection threshold.}

	\State{
		$D \gets \emptyset$,
		$G_c \gets \emptyset$,
		$D_{Gc} \gets \emptyset$,
		$attack \gets false$
	}
	\For{$x : x \in S$}
		\State{$c \gets F(x)$}
		\If {$G_c == \emptyset$} 	 \hfill \# sets and threshold initialization
			\State{$G_c \cup \{x\}$, $D_{Gc} \cup \{0\}$, $T_c \gets 0$}
		\Else
			\State{$d \gets \emptyset$}
			\ForAll{$y : y \in G_c$} \hfill \# pairwise distance
				\State{$d \cup \{dist(y, x)\}$}
			\EndFor
			\State{$d_{min} \gets min(d)$} \hfill \# distance to closest element
			\State{$D \cup \{d_{min}\}$} \hfill \# add distance to $D$
			\If {$d_{min} > T_c$} \hfill \# sets and threshold update
				\State{$G_c \cup \{x\}$}
				\State{$D_{Gc} \cup \{d_{min}\}$}
				\State{$T_c \gets max(T_c, \overline{D_{Gc}} - std(D_{Gc}))$}
			\EndIf
		\EndIf
		\If {$ \vert D \vert > 100$} \hfill \# analyze distribution for $D$
			\State{$D' \gets  \{z \in D, z \in \left\langle \overline{D} \pm 3 \times std(D) \right\rangle \}$}
			\If {$W(D') < \delta$} \hfill \# attack detection test
				\State{$attack \gets True$}
			\Else
				\State{$attack \gets False$}
			\EndIf
		\EndIf
	\EndFor
\end{algorithmic}
\end{algorithm}

%
%

Consider the stream $S$ of samples $x$ queried by a single client from the target model $F$. We calculate the minimum distance $d_{min}(x_i)$ between a new queried sample $x_i$ and any previous sample $x_{0,..,i-1}$ of the same class $c$. All $d_{min}(x_i)$ are stored in a set $D$.
By doing so, we want to model the distribution of distances between queried samples and identify samples abnormally close to or far from any previously queried sample.
For efficiency, we do not keep track of all past queries in $S$ but incrementally build a \textit{growing set} $G_c$ for each class $c$. $G_c$ consists only of samples whose distance $d_{min}$ is above a threshold value $T_c$.
We define $T_c$ as the mean minus standard deviation of the minimum distance $d_{min}$ between any two elements already in $G_c$.
The distance $d_{min}(x_i)$ is computed only w.r.t. elements in $G_c$ for $F(x_i) = c$.



Our attack detection criterion is based on quantifying how closely distances in $D$ fit a normal (Gaussian) distribution. We flag an attack if the distribution of these distances deviates too much from a normal distribution.
Figure~\ref{fig:query_distrib} illustrates the intuition why our approach can effectively detect model extraction. It depicts the difference in the distribution of distances (values in $D$) between benign and adversarial queries. We see that benign queries to the MNIST and GTSRB models fit a distribution that is close to normal. However, adversarial queries produce spikes on several values resulting in skewed distributions. These correspond to synthetic samples for which the distance to previous samples is artificially controlled by the adversary. Other distances occur more seldom and correspond mostly to natural seed samples queried at the beginning of the attacks. Such trends are typical for all known attacks.

Several metrics exist to quantify this phenomenon and evaluate if a set of values fits a normal distribution, i.e., to perform a \textit{normality test}. We considered and tested three, namely the Anderson-Darling test~\cite{anderson1952asymptotic}, the Shapiro-Wilk test~\cite{shapiro1965analysis} and the K-squared test~\cite{d1971omnibus}.
The Shapiro-Wilk test was selected because the values of its test statistic $W$ produced the largest difference when computed on benign and adversarial queries. A prior study also concluded that the Shapiro-Wilk test has the most predictive power to assess whether a set of values fits a normal distribution~\cite{razali2011power}.
The test statistic $W$ used in the Shapiro-Wilk test is given in Eq.~\ref{eq:shapiro}, where $x_{(i)}$ is the $i^{th}$ order statistic in the sample $D$, $\overline{x}$ is the sample mean, and $a_i$ are constants related to the expected values of the order statistics. More details are provided in~\cite{shapiro1965analysis}. $W$ is defined on $[0,1]$ and a low value highlight deviation from a normal distribution.

\begin{equation}
\label{eq:shapiro}
	W(D) =
	\frac{\left( \sum_{i=1}^n a_i x_{(i)} \right)^2 }
	{\sum_{i=1}^n (x_i - \overline{x})^2} \text{, for } D = \left\lbrace x_1, \ldots , x_n \right\rbrace
\end{equation}

Algorithm~\ref{algo:detection} describes \ourname's detection technique in detail.
The detection process starts when a client queries at least 100 samples ($|D| > 100$) because a sufficient number of values is necessary to compute a relevant $W$. We first remove outliers from $D$, i.e., values being more than 3 standard deviations away from the mean of values in $D$. According to the \textit{68-95-99.7} empirical rule, 99.7\% of values coming from a normal distribution belong to this interval.
We compute the Shapiro-Wilk test statistic $W$ on the resulting set deprived from outliers $D'$.
Next, if $W(D')$ is below a threshold $\delta$, \ourname detects an extraction attack.

\ourname requires the defender to set one parameter: the detection threshold $\delta$.
It also needs a domain-specific distance metric $dist()$ to compute distances between inputs. We use $L^2$ (Euclidean) norm for image samples in our experiments.

\subsection{Evaluation}
\label{sec:eval-detection}

We evaluate \ourname in terms of success and speed. Speed refers to the number of samples queried by an adversary until we detect the attack. It correlates with the amount of information extracted and must be minimized. We also evaluate the \emph{false positive rate} (FPR): the ratio of false alarms raised by our detection method to all query sequences from benign clients.

To evaluate success, we assess its detection of attacks against the two target models previously trained in Sect.~\ref{sec:exp-setup} for MNIST and GTSRB datasets.
We subject these models to four different attacks:
\tramer, \papernot and our new \textit{IFGSM} \trnd-64 (noted \trnd here) and \colorr-25 attack (noted \colorr here).
We use the samples generated while evaluating the performance of these attacks in Sect.~\ref{sec:eval} and query the prediction model with them one by one (in the order they were generated). \ourname's detection algorithm is computed for each new queried sample. When an attack is detected, we record the number of samples queried until then by the adversary to evaluate the speed of detection.

To evaluate the false positive rate, we use natural data from MNIST and GTSRB datasets. To demonstrate that \ourname is independent of a specific data distribution, we also use randomly generated samples (images with uniformly random pixel values), the U.S. Postal Service (USPS)~\cite{lecun1990handwritten} and Belgian traffic signs (BTS)~\cite{Timofte-WACV-2009} datasets. USPS and BTS datasets contain similar data as MNIST and GTSRB respectively but from different distributions. We reshaped the samples to fit the input size of MNIST and GTSRB models.
We simulate a benign client by randomly picking 6,000 samples from a given dataset and successively querying the appropriate model: \textit{MNIST/USPS/random} $\rightarrow$ MNIST model, \textit{GTSRB/BTS/random} $\rightarrow$ GTSRB model. We simulate five benign clients per dataset (30 clients).
To evaluate FPR, we split this sequence of queries into 120 chunks of 50 queries each and count a false positive if at least one alert is triggered by \ourname in a chunk.
Successive benign queries can also be related to each other.  
In a self-driving car scenario, successive pictures of the same road sign are taken and submitted to the model while getting closer to it. We simulated this scenario using the GTSRB validation set that contains 207 \textit{sequences} composed of 30 pictures each of a single road sign taken from a decreasing distance (6,210 samples). We ran five tests, randomly shuffling the order of sequences and submitting them to the GTSRB model while computing the FPR.

\begin{figure}[tbh]
	\begin{subfigure}{0.49\columnwidth}
	\centering
	\includegraphics[width=\columnwidth]{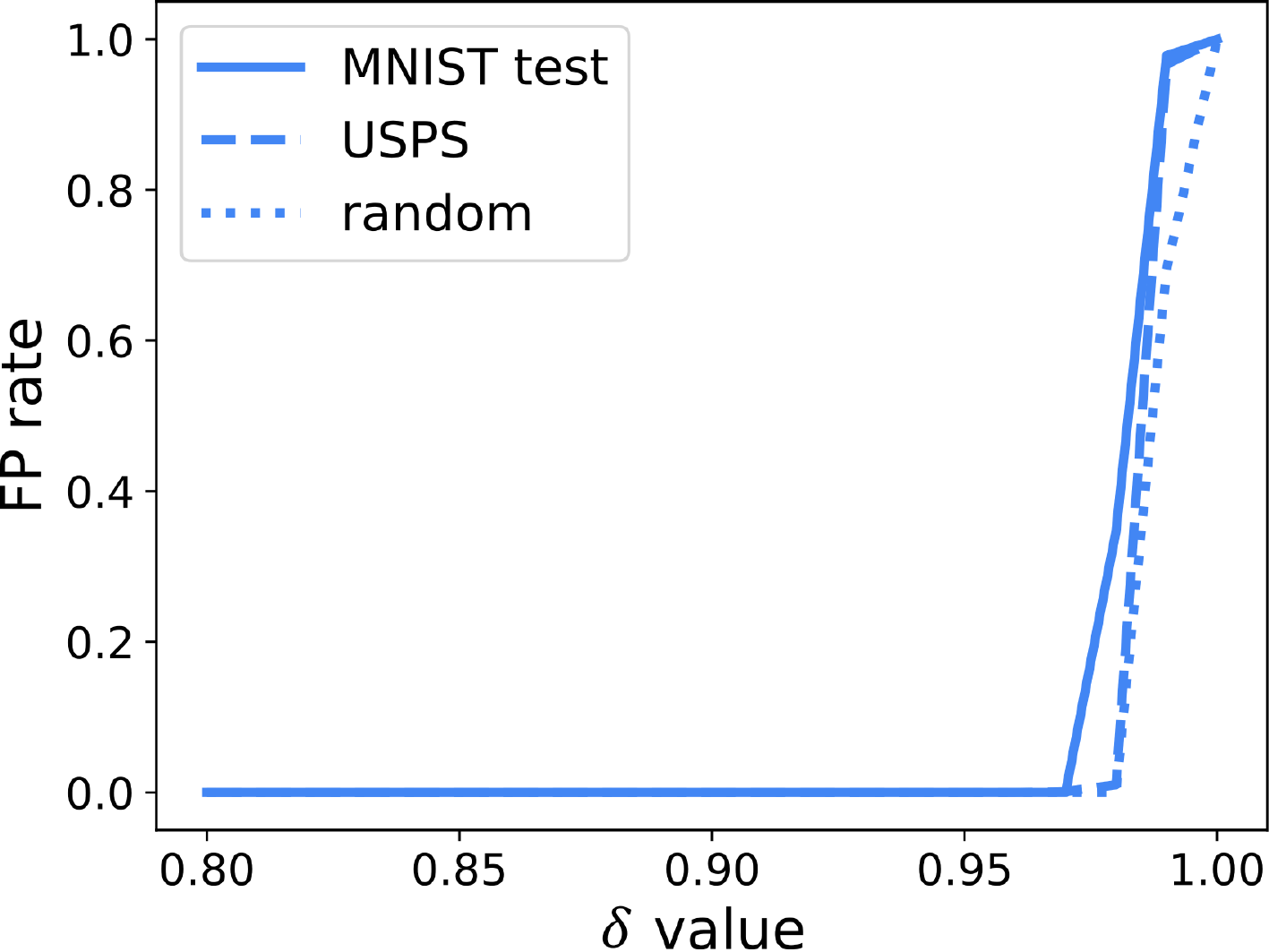}
		\captionof{figure}{MNIST model}
	\label{fig:fp_mnist}
	\end{subfigure}
	\begin{subfigure}{0.49\columnwidth}
	\centering
	\includegraphics[width=\columnwidth]{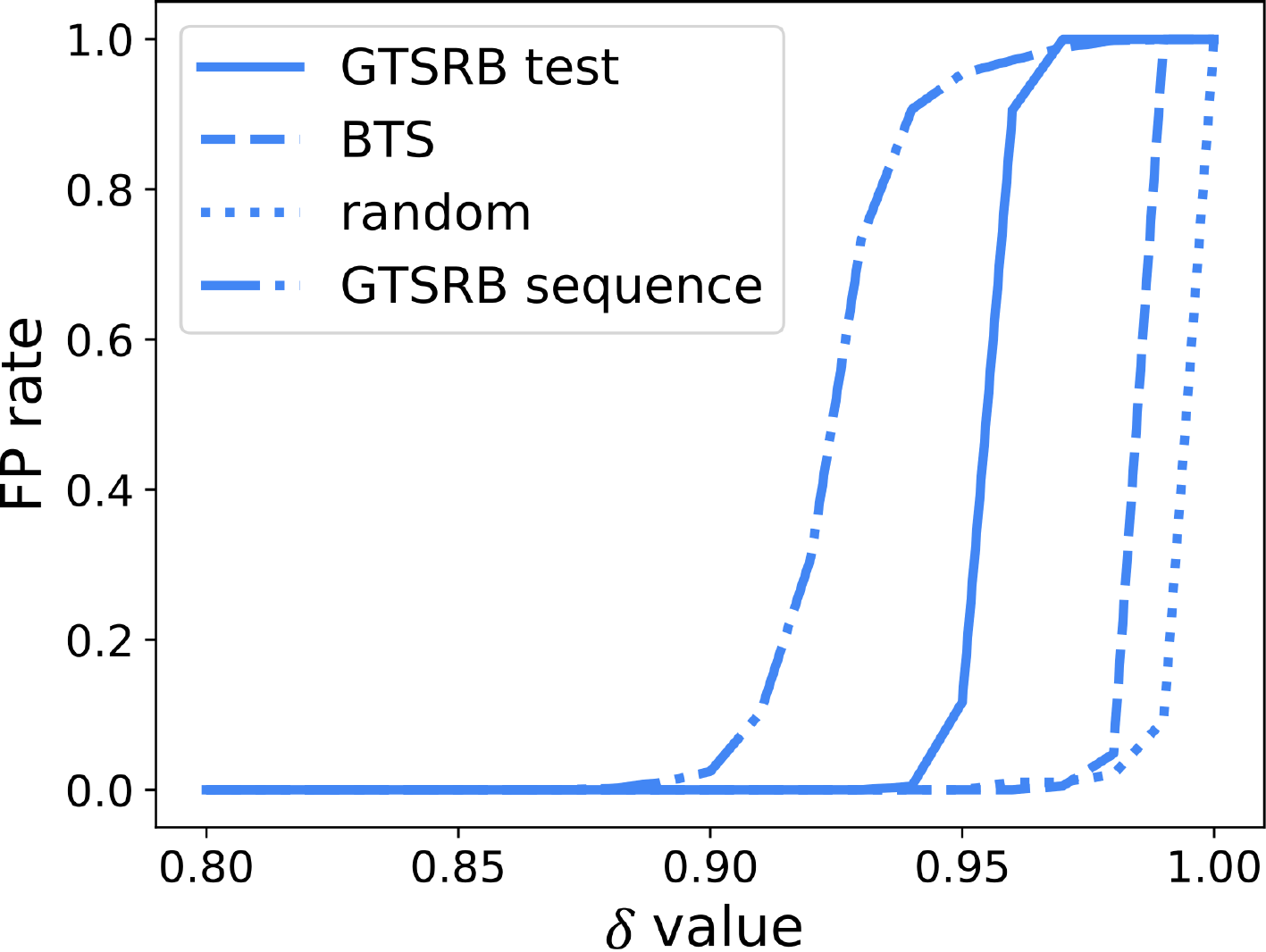}
		\captionof{figure}{GTSRB model}
	\label{fig:fp_gtsrb}
	\end{subfigure}
	\caption{\label{fig:fp_rate} FPR for \ourname vs. detection threshold $\delta$.}

\end{figure}

Figure~\ref{fig:fp_rate} depicts the increase in FPR according to the detection threshold value $\delta$.
A high $\delta$ value of $0.96$ results in no false positives for MNIST model while GTSRB model requires $\delta=0.87$ to reach the same result. It is worth noting that different simulated benign users generate queries with different distributions in the GTSRB experiment (Fig.~\ref{fig:fp_gtsrb}). While the \textit{BTS} and \textit{random} queries are distributed close to normal (high $\delta$ values result in no false positives), GTSRB queries are further away from a normal distribution (lower $\delta$ values required for no false positives). In the \textit{GTSRB sequence} queries, images in a sequence have a relatively small distance to each other while images from different sequences have higher distances between them. This explains a more scattered distribution of distances that deviates from a normal distribution. This shows that $\delta$ is a domain specific parameter that needs to be set with respect to the model to protect and its use case scenario.

Table~\ref{tab:queries} presents detailed speed of detection for a few selected $\delta$ values. Most attacks are detected shortly after they have a change in their query strategy. From natural to synthetic samples for \papernot, \trnd and \colorr attack (after 100 queries for MNIST and 430 queries for GTSRB). From random samples to line search strategy for \tramer attack (after 5,000 queries). While the detection is slower for the \tramer this is not a concern since it is itself slow in extracting DNNs. 
An estimate for the performance of the substitute model at the time of detection can be found in Tab.~\ref{table:summary} (no synthetic queries).
The \trnd attack is the only one that remains undetected against the GSTRB model if $\delta$ is too low. This is because the it uses a large step size $\lambda=64$ which produces synthetic samples with a large distance between each other. This distance happens to fit the normal distribution of natural samples for the GTSRB model.
By increasing $\delta$ to 0.94, \ourname effectively detects the \trnd attack while producing a few false positives (0.1\%). This $\delta$ value cannot be applied to all scenarios though, e.g., it triggers a large number of false positives with \textit{GTSRB sequence} queries (cf. Fig.~\ref{fig:fp_gtsrb}).

For most of our tests, this demonstrates that \ourname is effective at protecting against most model extraction attacks developed to date. Using an appropriate $\delta$ threshold, it detects quickly \tramer, \papernot and \colorr attacks while avoiding false positives for benign queries across the tested datasets: MNIST, USPS, GTSRB (+ sequence), BTS and random queries. A more careful selection of $\delta$ is necessary to detect the \trnd attack against GTSRB, and it may not apply to any model deployment scenario (e.g., high FPR in \textit{sequence} scenario), meaning that \ourname perhaps cannot be reliably deployed in all scenarios, as it may limit the usability.

\begin{table}
	\caption{Adversarial queries made until detection respect to $\delta$ value. FPR is averaged over all simulated benign query scenarios (* = \textit{GTSRB sequence} was discarded). \colorr attack is only performed against the GTSRB model so no results are reported for MNIST.}
	\label{tab:queries}
	\begin{center}
	\begin{tabular}{ c  c  c  c c c}
		\hline
		\multirow{2}{*}{\textbf{Model ($\delta$ value})} & \multirow{2}{*}{\textbf{FPR}} & \multicolumn{4}{c}{\textbf{Queries made until detection}} \\
		& & \tramer & \textsc{Pap.} & \trnd & \colorr  \\
		\hline
		\textbf{MNIST} ($0.95$) & 0.0\% & 5,560 & 120 & 140 & -\\
		\textbf{MNIST} ($0.96$) & 0.0\% & 5,560 & 120 & 130 & -\\
		\textbf{GTSRB} ($0.87$) & 0.0\% & 5,020 & 430 & missed & 550\\
		\textbf{GTSRB} ($0.90$) & 0.6\% & 5,020 & 430 & missed  & 480 \\
		\textbf{GTSRB} ($0.94$) & 0.1\%* & 5,020 & 430 & 440 & 440  \\
		\hline
	\end{tabular}
	\end{center}
\end{table}

To estimate the overhead of \ourname, we computed the memory required to store the growing set $G$.
Note that $G$ represents a subset of all queries $S$.
Samples for MNIST and GTSRB models have an average size of 561B and 9.3kB respectively. The average memory required before detecting  \papernot and  \textit{T-RND} attack for MNIST is around 55kB ($561B \times 98$ samples) and 3.2MB for GTSRB ($9.3kB \times 343$ samples).
Benign clients generate a larger $G$ since its growth is not stopped by a detection. However, this growth naturally slows down as a client makes more queries.
As an estimate, we used 1.9MB (MNIST test: $561B \times 3,374$ samples) and 1.0MB (USPS: $561B \times 1,804$ samples) for storing $G$ of a MNIST model client submitting 6,000 queries. We used 28.1MB (GTSRB test: $9.3kB \times 3,025$ samples) and 30.2MB (BTS: $9.3kB \times 3,254$ samples) for storing $G$ of a GTSRB model client submitting 6,000 queries.

\subsection{Discussion}
\label{sec:discussion}


\begin{figure}[ht]
		\begin{subfigure}{0.49\columnwidth}
	\centering
	\includegraphics[width=\columnwidth]{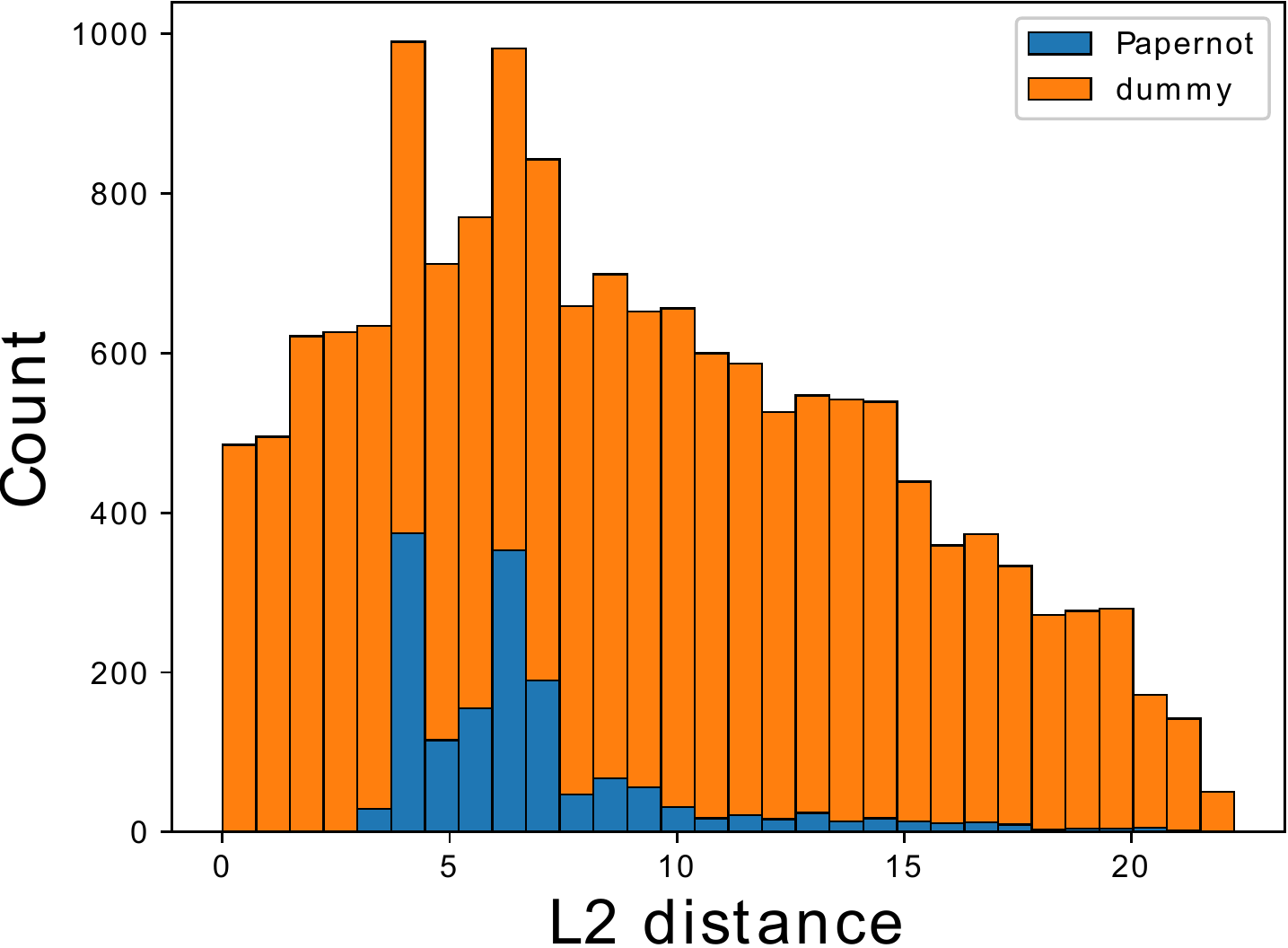}
		\captionof{figure}{MNIST model}
	\end{subfigure}
	\begin{subfigure}{0.49\columnwidth}
	\centering
	\includegraphics[width=\columnwidth]{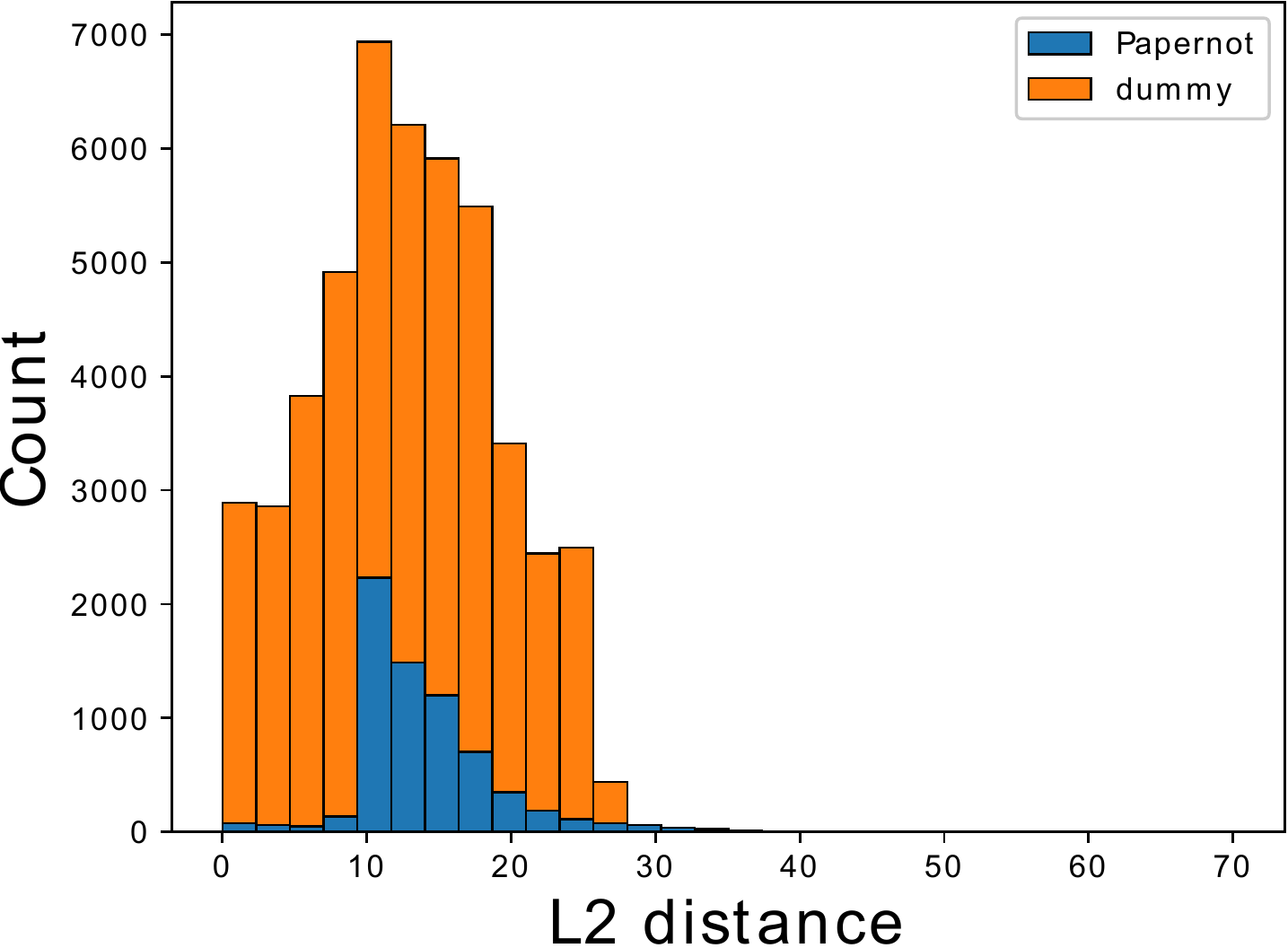}
		\captionof{figure}{GTSRB model}
	\end{subfigure}
	\caption{\label{fig:adaptive_distr} Example of controlling the distribution of $D$ under \papernot attack. \textbf{Left:} 1,600 attack queries (blue) against MNIST model. 14,274 dummy queries (orange) are required to avoid detection. \textbf{Right:} 6,680 attack queries against GTSRB model, 41,160 dummy queries are required to avoid detection.}

\end{figure}

\noindent\textbf{Evasion of Detection:}
We observed that \ourname can be evaded by \papernot and \trnd type of attacks by carefully selecting a step size $\lambda$ that would simulate a normal distribution of samples (cf. Sect.~\ref{sec:eval-detection} \trnd attack against GTSRB). We also concluded that $\lambda$ is a important factor impacting the success of a model stealing attack (cf. Sect~\ref{sec:synthetic-eval}). Evading \ourname by modifying the step size $\lambda$ optimal for model stealing purposes may degrade the performance of the stolen model. 

Alternatively, an adversary can attempt to evade detection by making \textit{dummy queries} that are not useful for building the substitute model but that would maintain a normal distribution of distances between queries. To evaluate the additional query cost of this evasion attack, we simulated an adaptive adversary.
Following Kerckhoffs's Principle, it has full knowledge of the detection algorithm, including the detection threshold value $\delta$ that is supposed to be secret.

In order to control the query distribution, a dummy query needs to satisfy two conditions: (1) it must be compared to a selected subset $G_c$ of the \textit{growing set} (targeted classification) and (2) its minimum distance $d_{min}$ (cf. Alg.~\ref{algo:detection}) must help to avoid detection, i.e., increase $W(D)$ to satisfy $W(D) \geq \delta$. Dummy queries must complement $D$ to fit a normal distribution as shown in Fig.~\ref{fig:adaptive_distr}.
Table~\ref{tab:adaptive} presents the query overhead required to circumvent \ourname. It ranges from $+3 \times$ to $+10 \times$ more queries depending on the target model and attack considered.

\begin{table}
	\caption{Increased query cost to circumvent \ourname (\textbf{GTSRB} \trnd* with $\delta = 0.94$).}
	\label{tab:adaptive}
	\begin{center}
	\begin{tabular}{lcccc}
		\hline
		\multicolumn{1}{r}{\textbf{Model}} & \multicolumn{3}{c}{\textbf{MNIST} ($\delta = 0.96$)} \\
		\textbf{Attack} & \papernot & \textit{T-RND} & \tramer \\
		\hline
		\textbf{Original queries}  & 1,600 & 1,600 & 10,000 \\
		\textbf{Additional queries}  & 14,274 & 4,764 & 79,980 \\
		\textbf{Overhead} & $+890\%$ & $+300\%$ & $+800\%$ \\
		\hline
	\end{tabular}
	\newline
	\vspace*{0.5 cm}
	\newline
	\begin{tabular}{lcccc}
		\hline
		\multicolumn{1}{r}{\textbf{Model}} & \multicolumn{4}{c}{\textbf{GTSRB} ($\delta = 0.90$)} \\
		\textbf{Attack} & \papernot & \trnd* & \tramer & \colorr \\
		\hline
		\textbf{Original queries}  & 6,680 & 6,680 & 10,000 & 6,680 \\
		\textbf{Additional queries}  & 41,160 & 19,986 & 99,990 & 39972 \\
		\textbf{Overhead} & $+620\%$ & $+300\%$ & $+1000\%$ & $+600\%$ \\
		\hline
	\end{tabular}
	\end{center}
\end{table}

In our evaluation, we controlled the distribution of $D$ using selected $d_{min}$ values and without generating queries. Thus our evaluation provides an estimated lower bound on the number of queries required to circumvent \ourname. We experimented with several strategies concluding that creation of such queries is not trivial. Notably, the following strategies did not fool our detection algorithm:
\begin{itemize}
	\item [1)] random noise drawn from normal/uniform distribution
	\item [2)] natural samples of desired class perturbed with random noise 
	\item [3)] like \textit{(1)} and \textit{(2)} but constraining $d_{min}$ to be in range $\left\langle \overline{D} \pm std(D) \right\rangle$ or $\left\langle \overline{D} \pm 2 \times std(D) \right\rangle$
	\item [4)] like \textit{(1)} and \textit{(2)} but submitting the sample only if it satisfies the $W(D) \geq \delta$.
\end{itemize}
In case of strategies \textit{(1)} and \textit{(2)} we observed that \textit{dummy queries} and \textit{useful queries} formed two spaced out peaks in the distribution. For \textit{(3)} the underlying seed samples impacted $d_{min}$ too much unless the noise was large enough to be equivalent to \textit{(1)} or \textit{(2)}. Finally, in \textit{(4)}, after several samples ($\approx 50$) the search for individual images became too time consuming (thousands of samples) to find a single valid query.


Since \ourname analyses samples queried by a single client, an adversary can distribute its queries among several clients to avoid detection (\textit{Sybil attack}). Using a sufficient number of clients, \ourname can be circumvented.

\noindent\textbf{Countermeasures:} Once \ourname detects an attack, we must resort to effective mitigation. Blocking requests from the adversary would be a straightforward prevention. This would be effective on single-client models protected by local isolation.
The defender might also deceive the adversary with altered predictions once an attack is detected in order to degrade the substitute model learned by the adversary.
Returning the second or third class with the highest likelihood according to the prediction of the target model may plausibly deceive the adversary into thinking it has crossed a class boundary while it has not and effectively undermine its substitute model.

\noindent\textbf{Generalizability:} \ourname is applicable to many types of data and ML models without any alterations since its design is independent from these considerations and only relies on identifying adversarial querying behavior.
The only aspect that depends on the type of data is finding a distance metric appropriate to compute differences between input samples of a certain type, e.g., we chose $L^2$ norm for image input. Alternatively, any $L^p$ norm or the structural similarity metric~\cite{sharif2018suitability} could be used for image input. As examples for other domains, the decibel metric (\textit{dB}) can be used on audio input~\cite{carlini2018audio} and the $L^1$ norm on malware input~\cite{grosse2017adversarial}.

One must also set an appropriate detection threshold $\delta$. This is dependent on the use case scenario for the model which will define a "benign" distribution of queries, as highlighted in Sect.~\ref{sec:eval-detection}. This value can be fixed using a training period during which only benign queries are submitted to the system and $\delta$ is selected as a maximum value that does not generate any false positives as we showed in Fig.~\ref{fig:fp_rate}.
Capturing this benign distribution correctly will impact the detection efficacy of \ourname.


\noindent\textbf{Storage overhead and scalability:} \ourname requires keeping track of several client queries, substantially increasing memory consumption. It is worth noting that we presented results for the extreme case of image classification models, which use high dimensional inputs. Nevertheless the amount of memory required per client was estimated to be a few megabytes (1-30 MB), which is reasonable. For local models being used by single clients, the storage requirements are thus minor.
Multi-client remote models serving up to a few hundred clients simultaneously will require a few gigabytes of memory in total. This is reasonable for a cloud-based setup where the model is hosted on a powerful server.



\section{Related Work}

\subsection{Model Extraction Attacks}
Model extraction is conceptually similar to concept learning~\cite{angluin1988queries,bshouty1996oracles} in which the goal is to learn a model for a concept using membership queries. Differences are that concepts to learn are not ML models and concept learning does not assume adversarial settings. Nevertheless, methods based on concept learning have been designed for adversarial machine learning and evading binary classifiers~\cite{lowd2005adversarial,nelson2012query}. Model evasion may be considered a theoretically simpler task than model extraction~\cite{stevens2013hardness}, and so far, the efficiency of model extraction attacks have not been demonstrated on DNNs.
The extraction of information from DNNs has been addressed in non-adversarial settings by compressing DNNs to simpler representations~\cite{bucilua2006model,hinton2015distilling} or by obtaining interpretable decisions from ML models~\cite{craven1996extracting,towell1993extracting}. These work do not apply to adversarial settings since they require white-box access to the target model and its training data.

We presented the two closest prior works to ours in Sect.~\ref{sec:prior}. 
Tramer et al.~\cite{tramer:2016:stealing} introduced several methods for extracting ML models exposed in online prediction APIs. They exploit the confidence values from predictions in a systematic equation solving approach to infer exact model parameters. In contrast to our work, this method addresses only the extraction of simple models such as logistic regression. This technique is ineffective at extracting DNN models (cf. Sect.~\ref{sec:synthetic-eval}).
Papernot et al.~\cite{Papernot:2017:PBA} introduced a method for extracting a substitute DNN model for the specific purpose of computing transferable non-targeted adversarial examples. Their main contribution is the JbDA technique for generating synthetic samples (cf. Sect.~\ref{sec:prior}). 

In contrast to these works, we introduce a generic method for extracting DNNs. It is multipurpose and has higher performance in transfer of targeted adversarial examples and reproduction of predictive behavior. 

Alternative model stealing attacks assume access to large sets of natural samples and use active learning strategies to select the best samples to query~\cite{DBLP:journals/corr/abs-1812-02766}. In this paper, we consider a different adversary model with limited access to natural samples.
A recent line of work targets the extraction of model hyperparameters and architecture. Joon et al.~\cite{joon2018towards} train a supervised classifier taking as input $n$ prediction values rendered by a classifier for a fixed set of $n$ reference samples. Using this technique, they infer with significant confidence the architecture, optimization method, training data split and size, etc. of a confidential target model. Hua et al.~\cite{hua2018reverse} target a model locally isolated with hardware security mechanisms (Intel SGX) and introduce a hardware side channel attack to infer similar model information.
Another work~\cite{stealing2018wang} takes a stronger adversary model (access to training data) and introduces a technique for computing the value for the hyperparameter for $L^2$-regularization. 
\cite{joon2018towards} is complementary to our attack and can be used in the first stage to select the architecture for the substitute model.

\subsection{Defenses against Model Extraction}
A first defense to model extraction is to reduce the amount of information given to an adversary by modifying the model prediction. 
Prediction probabilities can be quantized~\cite{tramer:2016:stealing} or perturbed to deceive to the adversary~\cite{lee2018defending}.
We have shown that model extraction attacks are effective even without using prediction probabilities (Sect.~\ref{sec:eval}), making this line of defenses ineffective.
A second line of defense consists in detecting model extraction attacks by recording requests from clients and computing the feature space explored by the aggregated requests~\cite{Kesarwani2017model}. When the explored space exceeds a pre-determined threshold, an extraction attack is detected. Quiring et al.~\cite{8406619} use the same intuition and study the closeness of queries to class boundaries to detect model stealing attacks.
These techniques have limitations since they require linearly separated prediction classes (both are applied to decision trees). Thus they do not apply to high dimensional input spaces nor to DNN models, which build highly non-linear decision boundary in this space. The false alarm rate of this technique is not evaluated and might be high since a legitimate client can genuinely explore large areas of the input space. 
On the contrary, \ourname applies to any input data dimensionality and any ML model. It is effective at detecting model extraction attacks developed to date and does not degrade the prediction service provided to benign clients.

Alternatively, methods for detecting adversarial examples can help detecting synthetically generated samples from \textit{Papernot} attack and ours. The main approaches rely on retraining the model with adversarial samples~\cite{tramer2017ensemble}, 
randomizing the decision process~\cite{feinman2017detecting} or analyzing the inputs distribution~\cite{Meng2017magnet}.
These techniques assume a specific distribution of the benign inputs to the prediction model, i.e., the same distribution as the training data. Consequently, they may raise a high number of false alarms if benign clients request natural samples distributed differently than the training data. 

In contrast, \ourname has been developed in mind of avoiding false positives. It does not assume any training data distribution but only studies the evolution in distribution of samples submitted by a given client. This explains why we have low or no false positives even when analyzing benign data from diverse distributions.
Methods for detecting adversarial examples may not generalize to detected the \textit{Tramer} class of attacks~\cite{tramer:2016:stealing} since it does not rely on methods for crafting adversarial examples. 


\section{Conclusion}

We have systematically explored approaches for model extraction. We evaluated several attacks on different DNN models and showed that hiding hyperparameters of the target model does not help protect against model extraction. 
Reducing DNN outputs from classification probabilities to labels only has
nearly no impact on prediction accuracy, but does impact transferability of adversarial examples.
Keeping model architectures confidential helps to protect against model extraction attack and transferable adversarial examples.  
In scenarios where it is possible, limiting the adversary's access to natural seed samples, can also limit the effectiveness of model extraction.

Recent research has shown that ML models, especially DNNs, suffer from various vulnerabilities. 
Consequently, protecting confidentiality of models is a useful mitigation. In this black-box scenario, an attacker is forced to repeated interactions with the model. We demonstrated that model extraction can be effectively detected by collecting \textit{stateful} information of queries in \textit{ML prediction APIs}. This defense has significant advantages since it does not require any knowledge about the ML model, nor about the data used to train it.
Relying on deviations from benign distributions, we found it can be circumvented 
if the attacker mimics such distributions. We leave robustness against such an attacker 
to future work. 
Model confidentiality combined with a stateful defense strategy is a promising venue for effectively protecting ML models against a large range of adversarial machine learning attacks. 
One example we are currently exploring is defending against black-box attacks for forging adversarial examples (without resorting to building substitute models via model stealing attacks; see Appendix). Such attacks usually require thousands of queries to forge one adversarial example. A stateful prediction API like the one described in this paper with \ourname appears to be a promising defense direction.

\section{Acknowledgments}
This  work  was  supported  in  part  by  the  Intel  Collaborative  Institute  for  Collaborative  Autonomous  and  Resilient Systems (ICRI-CARS) and by the SELIoT project and the Academy of Finland under the WiFiUS program (grant 309994). 
We thank Alexey Dmitrenko and Buse Gul Atli for their help in evaluating our work.


%
\bibliographystyle{IEEEtranS}
\bibliography{main}  

\appendix

\section{Adversarial examples and black box attacks}
\label{app:adversarial-examples}



We defined \emph{black-box} adversaries with \emph{surrogate data} in our paper. 
In addition, the following \emph{black-box} adversarial attacks have been examined in literature:

\paragraph{Surrogate Learner} This setting is similar to ours, in that a substitute
model is used for the adversarial attacks. However, Munoz et al.~\cite{munoz2017towards} state that
this threat model does not assume knowledge on what type of target classifier is
used, but may use \emph{same training data}. 
Adversarial examples for ImageNet models~\cite{krizhevsky2012imagenet} are typically
shown in this setting, e.g. Dong et al.~\cite{dong2017boosting} show it is possible
to create highly transferable adversarial examples. 


\paragraph{Finite difference methods} It is also possible to create targeted adversarial examples for DNNs without substitute models~\cite{chen2017zoo,ilyas2017query}. These attacks are very effective, but have limitations: these attacks require \emph{thousands of queries per sample} and may be therefore easily detectable, they do not extract models and mostly require access to target model probabilities. We calculated that attacking MNIST with Natural Evolution Strategies~\cite{ilyas2017query} requires on average several 1000s queries on MNIST, and several 100s queries on GTSRB per adversarial example.

\end{document}